\documentclass{article}
\usepackage[utf8]{inputenc}
\usepackage[margin=1in]{geometry}
\usepackage{amsmath}
\usepackage{graphicx}
\usepackage{caption}
\usepackage{subcaption}
\usepackage{bm}
\usepackage{amsfonts} 

\DeclareMathOperator*{\argmin}{arg\,min}
\usepackage{dsfont}
\usepackage{placeins}
\usepackage{booktabs}
\usepackage{siunitx}
\usepackage[table]{xcolor}
\usepackage{hyperref}
\usepackage{fancyhdr}

\usepackage{authblk}
\usepackage{mdframed}

\usepackage{multirow}

\newcommand\tophline{\hline\noalign{\vspace{1mm}}}
\newcommand\middlehline{\noalign{\vspace{1mm}}\hline\noalign{\vspace{1mm}}}
\newcommand\bottomhline{\noalign{\vspace{1mm}}\hline}
\usepackage{makecell}
\usepackage{stmaryrd}

\usepackage{natbib}
\usepackage{doi}
\title{A barycenter-based approach for the multi-model ensembling of subseasonal forecasts}
\author[1]{Camille Le Coz}
\author[1]{Alexis Tantet}
\author[2]{Rémi Flamary}
\author[1]{Riwal Plougonven}
\affil[1]{LMD/IPSL, École Polytechnique, Institut Polytechnique de Paris, ENS, PSL Research University, Sorbonne Université, CNRS, Palaiseau France}
\affil[2]{CMAP, Ecole Polytechnique, Institut Polytechnique de Paris, Palaiseau, France}
\date{}

\begin{document}

\maketitle

\begin{abstract}
Ensemble forecasts and their combination are examined from the perspective of probability spaces. By treating ensemble forecasts as discrete probability distributions, multi-model ensemble (MME) forecasts are reformulated as barycenters of these distributions. We consider two barycenters, each defined with respect to a different distance metric: the $L_2$ barycenter, which corresponds to the traditional pooling method, and the Wasserstein barycenter, which better preserves certain geometric properties of the input ensemble distributions. 
\newline
As a proof of concept, we apply the $L_2$ and Wasserstein barycenters to the combination of four models from the Subseasonal to Seasonal (S2S) prediction project database. 
Their performance is evaluated for the prediction of weekly 2 m temperature, 10 m wind speed, and 500 hPa geopotential height over European winters.  
By construction, both barycenter-based MMEs have the same ensemble mean but differ in their representation of forecast uncertainty. In particular, the $L_2$ barycenter has a larger ensemble spread, making it more prone to underconfidence. Although both methods perform similarly on average in terms of the Continuous Ranked Probability Score (CRPS), the Wasserstein barycenter performs better more frequently.
\end{abstract}

\thispagestyle{fancy}
\newpage
\section{Introduction}  
\subsection*{Multi-model ensemble methods (MME)}

Multi-model ensemble (MME) methods have been shown to improve forecast skill for different time scales, from short-range \citep{Heizenreder2006,Casanova2009} and medium-range \citep{Hamill2012,Hagedorn2012} to seasonal forecasting \citep{Palmer2004,Alessandri2011,Kirtman2014}. The added value of MME over single-model ensemble (SME) forecasts has been attributed to several factors. First, there is in general no ``best'' SME \citep{Hagedorn2005}, the relative performance of the SMEs varies depending on the target considered(i.e. region and variable of interest, metrics, etc.). The MME can take advantage of the complementary skills of the SMEs, thereby performing better on average. Second, \citet{Hagedorn2005} identified error cancellation and non-linearity of the metric as the main reasons for the MME's performance exceeding the average performance of SMEs. 
Third, MMEs allow us to explore a new dimension of uncertainty that remains unexplored by SMEs. Combining different models enables us to account for uncertainty due to {\it model formulation}: by construction, an SME takes into account uncertainties in model initialization and can introduce variations in parameterizations to sample some of the uncertainty due to parameterization choices, but it cannot account for uncertainty due to model formulation. 
In addition, \citet{Weigel2008} showed that MMEs can improve predictive skill if and only if the SMEs ``fail to capture the full amount of forecast uncertainty''. 
Finally, MMEs benefit from their larger ensemble size, which often implies a better sampling of the underlying probability distribution.

\subsection*{MME for subseasonal to seasonal (S2S) forecasts}

Subseasonal to seasonal forecasting bridges the gap between weather (medium-range) and seasonal forecasting. It corresponds to the time range from two weeks up to two months. Predictions on this time scale are the focus of the Subseasonal to Seasonal (S2S) prediction project, whose objectives are to  improve forecast skills and promote their use \citep{Vitart2017}. 
As part of this project, a database containing S2S forecasts from thirteen (originally eleven) operational centers has been made available to the research community.
One of the main research questions the S2S project aims to address is: ``What is the benefit of a multi-model forecast for subseasonal to seasonal prediction and how can it be constructed and implemented?'' \citep{Vitart2017}.

Several studies have investigated the potential benefits of MMEs for S2S forecasts \citep{Vigaud2017a,Vigaud2020,Specq2020,Wang2020,Materia2020,Zheng2019,Pegion2019}. 
These studies use different MME methods, variables, and evaluation criteria, but they all concur that MMEs generally perform as well as, or better than, SMEs. Moreover, \citet{Vigaud2017a} and \citet{Specq2020} suggest that MMEs improve not only the skill but also the reliability of probabilistic forecasts. However, studies using the pooling method (i.e. the concatenation of ensembles) point out that the improved performance of MMEs is partly related to their larger number of members \citep{Zheng2019,Specq2020,Wang2020}. 
Among other studies, \citet{Karpechko2018} investigates a specific Sudden Stratospheric Warming event with a MME, while \citet{Ferrone2017} evaluates several MME methods, but neither compares the skill of MMEs to that of SMEs. Thus, while the potential benefits of MMEs at the subseasonal scale have been clearly established, the question of how best to combine ensemble forecasts from different models has received relatively little attention.  
The present study focuses on this question.

The most direct and commonly used method for multi-model combination is the ``pooling method'', which simply concatenates the ensemble members from the different models. The members of the resulting multi-model ensemble can either be equally weighted or be assigned different weights based on the model skills (e.g.~\citet{Weigel2008} for seasonal scale and \citet{Wanders2016} for subseasonal scale). In the aforementioned studies, all use some variation of this method except \citet{Vigaud2017a,Vigaud2020} and \citet{Ferrone2017}. The latter focuses on the prediction of terciles, and therefore couple multi-model combination with other post-processing methods (e.g.~model output statistics method such as extended logistic regression) to directly predict the terciles. 
Also focusing on quantiles, \citet{Gonzales2021} uses a sequential learning algorithm to linearly combine predictors (from the SMEs, as well as from climatology and persistence), with weights updated at each step based on previous performance. 
Other methods for weighted-MME have been explored, but have not yet been applied at the subseasonal scale. At the seasonal scale, \citet{Rajagopalan2002} developed a Bayesian methodology to combine ensemble forecasts for categorical predictands (also used by \citet{Robertson2004} and \citet{Weigel2008} at the seasonal scale to obtain terciles of the MME).
Other approaches, such as the Ensemble Model Output Statistics (EMOS) and Bayesian Model Averaging (BMA), adopt a probabilistic perspective and aim to construct one probability density function (PDF) from the different ensemble forecasts. The EMOS method assumes a specific parametric form for this predictive PDF, the parameters being obtained from the input ensembles by optimizing with respect to a chosen score (e.g.~the CRPS) over a training period \citep{Gneiting2005}. In contrast, the BMA assumes separate parametric forms for the PDFs of the input models (based on the ensemble forecasts). The predictive PDF is then a weighted mixture of these input PDFs, with weights given by the posterior probabilities of the input models \citep{Raftery2005}. 
The following framework of barycenter encompasses the BMA as will be discussed later.

\subsection*{MME as barycenters of discrete distributions}

In this study, we propose to revisit the combination of multiple ensemble models from a different perspective. We consider each ensemble forecast as a discrete probability distribution and reformulate the multi-model ensemble as the barycenter of these distributions. The barycenter then serves as a tool to merge several ensemble distributions into a single transformed ensemble distribution. 
We show that, in this framework, the pooled MME is actually the (weighted) barycenter with respect to the $L_2$ distance. Since we work with distributions instead of a collection of ensemble members, the notion of barycenter can be extended to other metrics in the space of probability distributions. In particular, a natural distance in this space is the Wasserstein distance that stems from optimal transport theory \citep{Villani2003}. The Wasserstein distance is defined as the cost of the optimal transport between two distributions.

Optimal transport and the Wasserstein distance have been used in diverse applications in climate and weather. 
They have been used to measure the response of climate attractors to different forcings \citep{Robin2017}, to evaluate the performance of different climate models \citep{Vissio2020}, or to assess different parameterizations \citep{Vissio2018}. Moreover, \citet{Papayiannis2018} used Wasserstein barycenters to point-downscale wind speed from an atmospheric model. 
\citet{Robin2019} developed a multivariate bias correction method based on optimal transport, while \citet{Ning2014} used it within the framework of data assimilation to deal with structural error in forecasts.

Here, we use the Wasserstein barycenter, or more precisely its Gaussian approximation, as a tool to build multi-model ensembles and compare it to the more traditional pooling method. The Wasserstein distance has some interesting properties in probability distribution space. While the $L_2$ distance measures the `` vertical'' distance between distributions, the Wasserstein distance is based on the ``horizontal'' displacement between them. This allows the Wasserstein barycenter to retain some geometric properties of the input distributions (such as normality, for example, see \citet{Backhoff2022}). 
We investigate the impact of changing the metric (from $L_2$ to Wasserstein) on the MME performance by applying both barycenters to the combination of four models from the S2S database.

The paper is organized as follows. The use of the two barycenters as MME methods is presented in Section~\ref{sec:MME}. We first link the pooling method to the $L_2$-barycenter (\ref{ssec:L2_bary}) before introducing the Wasserstein distance and its barycenter (\ref{ssec:ot_bary}). We apply these methods to the combination of four models from the S2S database \citep[namely, ECMWF, NCEP, ECCC, and KMA;][]{Vitart2017}. 
The case study is described in Section~\ref{sec:case_study}, including the datasets and evaluation metrics. The skill of the two MMEs and four SMEs is evaluated and compared in Section~\ref{sec:results}. The results are discussed in Section~\ref{sec:discussion}, and the main conclusions are presented in Section~\ref{sec:conclusion}.


\section{Multi-model ensemble methods \label{sec:MME}}

\subsection{Ensemble forecasts as discrete probability distributions \label{ssec:ens_as_dist}}
At the S2S time scale, it is necessary to move from a deterministic to a probabilistic approach using ensemble forecasting (e.g.~\citet{Kalnay2003}). 
The ensemble is typically generated by perturbing the initial conditions and running the model for each of these perturbed states. The set of perturbed initial conditions represents the initial uncertainty associated with possible errors in the initial state of the atmosphere. This initial uncertainty is then propagated forward in time by the model. Thus, an ensemble forecast is a set of $N$ perturbed forecasts representing the evolution in time of the probability density of atmospheric variables according to the model formulation. 

An ensemble forecast aims to sample the probability distribution of the forecasted variable conditioned on the initial distribution. Now, this can also be considered as a discrete probability distribution $\mu$ on the state space $\mathbb{R}^{n_t}$ such that $$\mu = \sum\limits_{i=1}^{N} a_i \delta_{\mathbf{x}_{i}}$$
where $N$ is the number of members in the ensemble, $\mathbf{x}_{i}\in\mathbb{R}^{n_t}$ is the position of the Dirac corresponding to the time series of the $i$-th member, $n_t$ is the number of lead times, and $a_i$ is the weight of the $i$-th member (such that for all $i\leq N , a_i\geq0$ and $\sum_{i=1}^{N}a_i=1$).
Thus, here,  $\mathbf{x}_{i}$ does not represent an instantaneous state, but a full time series. That is, $\mu$ is a multivariate distribution where the $n_t$ variables correspond to the forecasted values at different lead-times. 
In a standard ensemble forecast, all the members are equiprobable, so they have equal weights $a_i=\frac{1}{N}$. In the remainder of this section, we will look at ensemble forecast time-series from the perspective of discrete distributions. 

\subsection{Barycenters for multi-model combination \label{ssec:bary_mme}}
The goal of multi-model ensemble methods can be rephrased as combining the complementary information from these distributions to obtain a new discrete distribution that better represents the true probability distribution function of the forecasted variable. A way to summarize a collection of distributions  $(\mu_1, \mu_2, ..., \mu_K)$ is to compute their barycenter. The barycenter is found by solving the following minimization problem
\begin{align}
\argmin_{\mu}  \quad\sum_{k=1}^{K}\lambda_k . d(\mu_k,\mu)^2  \label{eq:barycenter}
\end{align}
where $\lambda_k$ represents the weights given to the distributions, and $d(\cdot,\cdot)$ is a distance between distributions. The barycenter, also known as the Fréchet mean, is effectively the distribution that best represents the input distributions with respect to a criterion given by the chosen distance $d$.

\subsubsection{$L_2$ barycenter\label{ssec:L2_bary}}

The $L_2$ distance between two distributions $\mu_1$ and $\mu_2$ is given by 
$ \left\Vert \mu_1-\mu_2  \right\Vert_2  = \left( \int_{\mathbb{R}^{n_t}} \left(\mu_1(\mathbf{z}) -
\mu_2(\mathbf{z})\right)^2 d^{n_t}\mathbf{z}\right)^{1/2} $. When using this distance in the barycenter
equation~\eqref{eq:barycenter}, one can find the analytical formula for the
$L_2$ barycenter of the distributions $\mu_1,...,\mu_K$:
\begin{align}
    \mu_{L_2}^\mathbf{\lambda} 
    & = \sum_{k=1}^{K}\lambda_k  \mu_k \label{eq:l2_bary} \\
    & = \sum_{k=1}^{K} \lambda_k  \sum_{i=1}^{N_k} a_{k,i} \delta_{\mathbf{x}_{k,i}}   \label{eq:l2_bary2}
\end{align}
where the distribution $\mu_k$ has $N_k$ Dirac masses located at points $(\mathbf{x}_{k,1},\ldots,\mathbf{x}_{k,N_k}) \in \mathbb{R}^{n_t,N_k}$ with associated probability vector $\mathbf{a}_k =(a_{k,1},\ldots,a_{k,N_k})=(\frac{1}{N_k},\ldots,\frac{1}{N_k})$ (for $k=1,\ldots,K$). 

Equation~\eqref{eq:l2_bary} shows that the $L_2$ barycenter is a weighted average of distributions. This is similar to the BMA method, which can then be seen as a $L_2$ barycenter with specific weights derived from the Bayesian framework (in which the weights are the posterior probabilities of the input models, i.e., they represent the probability that the associated model is the best model). However, Equation~\eqref{eq:l2_bary2} differs from BMA in the particular distributions being summed: an assumption on their shape is made in the BMA, while the ensemble distributions are directly used in the $L_2$ barycenter \citep{Raftery2005}.

For a generic choice of weights $\lambda_k$, one can see that the $L_2$ barycenter corresponds to the concatenation of the members of the $K$ input ensembles with the rescaling of the associated member weights (see Eq.~\eqref{eq:l2_bary2}). This is equivalent to ``pooling'' together the ensemble members from the different models. The pooling method is a simple and well-established MME method. 
It has been used at the seasonal scale \citep{Hagedorn2005,Weigel2008,Robertson2004,Becker2014}, at the decadal scale \citep{Smith2013} and at the medium-range scale \citep{Hamill2012,Hagedorn2012}. More recently, it has also been applied to subseasonal ensemble forecasts by \citet{Karpechko2018,Pegion2019,Zheng2019,Specq2020,Materia2020}. 
The pooling method is based on the assumption that all members from all ensembles are independent and identically distributed samples of the same distribution. Under this assumption, the members can be pooled to form an empirical distribution. 
However, this hypothesis is rarely verified in practice \citep{Kioutsioukis2014,Knutti2017}. 
If we assume instead that the input ensembles have different distributions because of model-specific errors (which we aim to sample), then interpreting the pooling method as a $L_2$ barycenter allows us to extend it to different distances. In the following, we introduce another distance, the Wasserstein distance, and its associated barycenter. 
From this point on, we consistently use $L_2$ barycenter to refer to pooling in order to facilitate comparison with this second barycenter-based MME method.

\subsubsection{Wasserstein barycenter \label{ssec:ot_bary}}

The Wasserstein distance stems from optimal transport theory and can be seen
as the cost of transportation between two distributions $\mu_1$ and $\mu_2$. It
can be defined on discrete distributions as: 
\begin{align}
    W_2^2(\mu_1,\mu_2) &= \ \min_{\mathbf{T}\in U(\mathbf{a}_1,\mathbf{a}_2)} \langle \mathbf{T}, \mathbf{C} \rangle  =   \min_{\mathbf{T}\in U(\mathbf{a}_1,\mathbf{a}_2)} \sum\limits_{i,j}t_{i,j}\|\mathbf{x}_{1,i}-\mathbf{x}_{2,j}\|^2  \label{eq:W2}
\end{align}
where $U(\mathbf{a}_1,\mathbf{a}_2)=\left\{ \mathbf{T}\in\mathbb{R}_{+}^{N_1,N_2}:
\mathbf{T}\mathbf{1}_{N_2}=\mathbf{a_1} ~~\text{and}~~
\mathbf{T}^T\mathbf{1}_{N_1}=\mathbf{a_2} \right\}$ is the set of all feasible
transport matrices $\mathbf{T}=\left(t_{i,j}\right)_{1\leq i\leq N_1, 1\leq j\leq N_2}$ between the probability vectors $\mathbf{a}_1$ and
$\mathbf{a}_2$ associated with $\mu_1$ and $\mu_2$ respectively (with $\mathbf{1}_{N}$ standing for the all-ones vector of size
$N$), and $\mathbf{C}\in\mathbb{R}^{N_1\times N_2}$ is a distance matrix whose
elements are the pairwise squared Euclidean distances between the elements of
$\mathbf{x}_{1,:}$ and~$\mathbf{x}_{2,:}$. Here, $\left( \langle \mathbf{T}, \mathbf{C} \rangle
\right)^{1/2}$ is the cost associated with the transport $\mathbf{T}$ (with $\langle . , . \rangle$ being the Frobenius inner product between matrices). The elements $t_{i,j}$ of a transport matrix~$\mathbf{T}$ describe the amount of
mass going from $\mathbf{x}_{1,i}$ to $\mathbf{x}_{2,j}$, while $\|\mathbf{x}_{1,i}-\mathbf{x}_{2,j}\|^2$ is the cost of moving one unit of mass from $\mathbf{x}_{1,i}$ to $\mathbf{x}_{2,j}$. The minimization problem consists of searching for the optimal transport among all the feasible ones in
$U(\mathbf{a},\mathbf{b})$, i.e., the transport associated with the lowest cost
denoted as the 2-Wasserstein distance. This is a short description of the Wasserstein
distance for discrete distributions; for more information see \citet{Peyre2018}, \citet{Santambrogio2015} or \citet{Villani2003}. 

Contrary to the $L_2$ barycenter above, there is no general closed-form expression for the Wasserstein barycenter, and one must solve the minimization problem~\eqref{eq:barycenter} to find it. 
However, computing the Wasserstein barycenter directly on discrete distributions 
causes an artificial decrease of the barycenter's variance, which leads to important under-dispersion of the multi-model ensembles. This variance shrinkage, illustrated in appendix~A, is due to the limited sample size of these distributions with respect to the number of dimensions (corresponding here to the number of lead times, see Section~\ref{ssec:ens_as_dist}).

\begin{figure}[!ht]
\centering
\includegraphics[width=\textwidth]{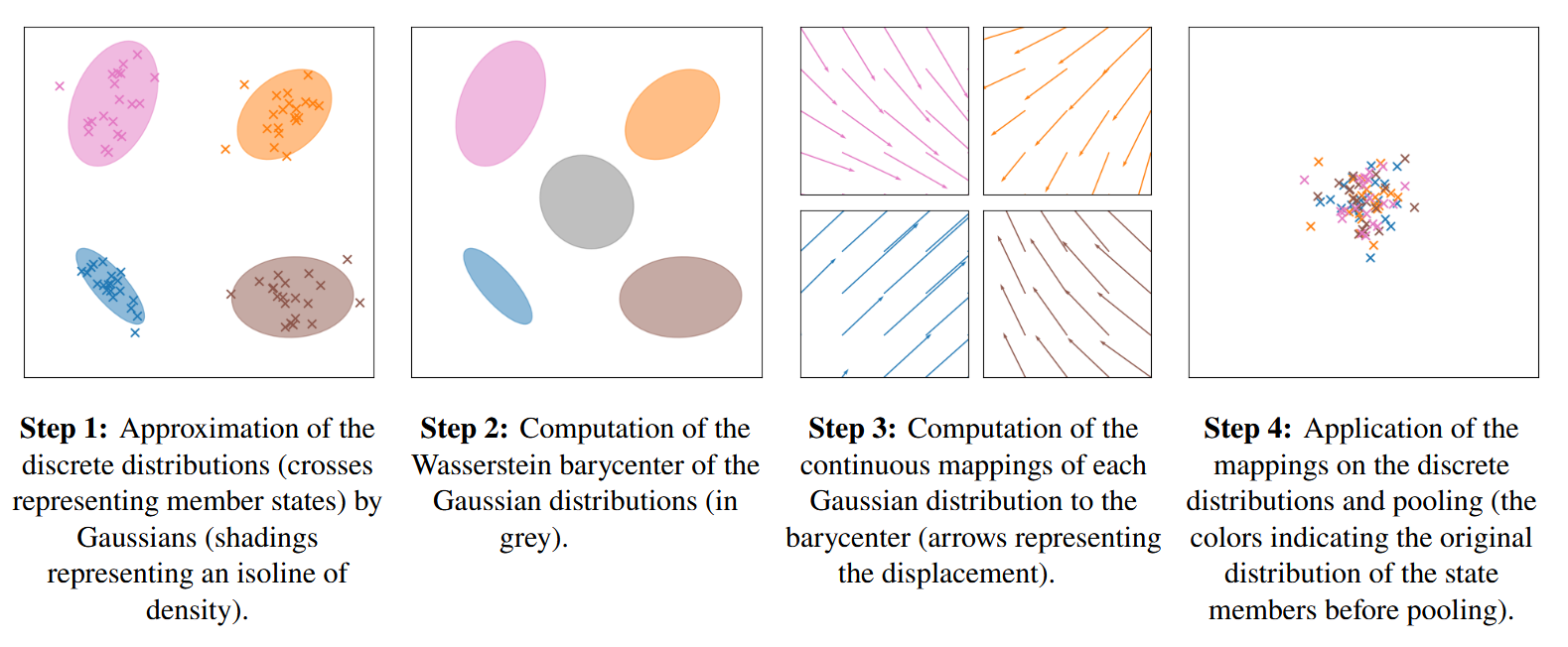}
\caption{Illustration of the Wasserstein barycenter with Gaussian mapping approach.}
\label{fig:GaussW2-bary}
\end{figure}

In order to avoid the variance shrinkage and have a more efficient and better statistical estimation, we propose to estimate the barycenter using a Gaussian approximation of the data and then to ``map'' all  forecast onto this barycenter using the Gaussian mapping.
That is, a Gaussian assumption is made to estimate a Gaussian Wasserstein barycenter which provides a closed-form affine mapping from each distribution to the barycenter, and then this mapping is applied to the discrete distributions which aligns them on the barycenter but keeps their individual samples. This means that only the first- and second-order moments of the Gaussian are estimated from the data which attenuates the statistical difficulty of estimating the Wasserstein distance \citep{flamary2020}. This approach was first used in \citet{Gnassounou2023} for data normalization of biomedical signals. Note that they additionally assumed the signal to be stationary and periodic to simplify the computations, however, these hypotheses are not appropriate to our dataset. 
The method is illustrated in Figure~\ref{fig:GaussW2-bary} and is made up of four steps: 
\begin{enumerate}
    \item Each discrete distribution $\mu_k$ is approximated by a multivariate normal distribution $\hat{\mu}_k$ of mean $\hat{\mathbf{m}}_k$ and covariance $\hat{\mathbf{\Sigma}}_k$ (for $k=1,\ldots,K$), which are estimated by the unbiased sample mean and covariance respectively.
    
      Note: in this study, the multivariate distribution does not represent the joint distribution between several meteorological variables, but the joint distribution between different time steps of the same meteorological variable (see Section~\ref{ssec:ens_as_dist}).
    
    \item The Wasserstein barycenter between the normal distributions is computed. It is also a  multivariate normal distribution with mean $\hat{\mathbf{m}}_b = \sum_{k=1}^K \lambda_k \hat{\mathbf{m}}_k$ and covariance $\hat{\mathbf{\Sigma}}_b$ that is estimated iteratively using the fixed-point iteration $\hat{\mathbf{\Sigma}}_b = \sum_{k=1}^K \lambda_k \left(\hat{\mathbf{\Sigma}}_b^{1/2}\hat{\mathbf{\Sigma}}_k^{1/2}\hat{\mathbf{\Sigma}}_b^{1/2}\right)^{1/2}$ from \cite{Agueh2011}. 

    \item The optimal transports from each input Gaussian distribution $\hat{\mu}_k$ to the Gaussian barycenter $\hat{\mu}_b$ are derived. 
    They are given by the following affine mappings  $T_k(\mathbf{x})= \mathbf{A}_k (\mathbf{x}-\hat{\mathbf{m}}_k) +\hat{\mathbf{m}}_b$  with $\mathbf{A}_k=\hat{\mathbf{\Sigma}}_k^{-1/2} \left(\hat{\mathbf{\Sigma}}_k^{1/2}~\hat{\mathbf{\Sigma}}_b~\hat{\mathbf{\Sigma}}_k^{1/2} \right)^{1/2} \hat{\mathbf{\Sigma}}_k^{-1/2}$ (assuming that the $~\hat{\mathbf{\Sigma}}_k$ are invertible, see below).     
    
    \item The mappings are applied to the corresponding discrete distributions (that is the mapping $T_k$ is applied to each state member $\mathbf{x}_{k,i}$ of $\mu_k$), and then the adapted distributions are pooled together (as for the $L_2$ barycenter in eq.~\eqref{eq:l2_bary2}) to create the 2-Wasserstein barycenter with Gaussian mapping ($GaussW_2$ barycenter hereafter).
\end{enumerate}

This approach computes mappings and barycenters on continuous (multivariate normal) distributions, thereby avoiding variance shrinkage of the barycenters due to limited discrete sample sizes (see Appendix~A). 
Computing the mean and covariance of a discrete distribution for the multivariate normal approximation is less sensitive to the sampling of the distributions \citep{Gnassounou2023}, and regularization can be used if needed to ensure the invertibility of the covariance matrices $\hat{\mathbf{\Sigma}}_k$. This is especially important  when the number of members (i.e. samples) is small compared to the number of dimensions.  
The approximation of the discrete distributions by multivariate normal distributions is a strong hypothesis considering that ensembles in climate science are propagated according to nonlinear evolution equations. However, this is a classical hypothesis for modeling stochastic processes and in signal processing. Moreover, this hypothesis is only used to derive Gaussian barycenter and the mappings which are then applied to the original discrete distributions which can be seen as a second-order calibration of the models onto a barycenter which preserves higher-order moments through the individual samples. 
For variables with a strongly non-Gaussian distribution (such as precipitation), it is also possible to first apply a transformation to make the distribution more Gaussian (e.g., using Gaussian anamorphosis; see \citet{Bertino2003, Lussana2021}), compute the barycenter in the transformed space, and then apply the inverse transform. 

Another advantage of this approach is the computational cost (in low dimensions). Estimating the parameters of the multivariate normal distribution from the discrete one has a computational complexity of $\mathcal{O}(N d^2)$ (where $N$ is the number of points in the discrete distribution, and $d$ the number of dimensions, here $d=n_t$ lead times). The Gaussian mappings and Gaussian barycenter computation have a complexity of $\mathcal{O}(d^3)$ and $\mathcal{O}(d^3 log(1/\epsilon))$ respectively (where $\epsilon$ is the tolerance in the fixed-point algorithm). On the other hand, computing the exact barycenter requires multiple resolutions (per model and iteration) of exact optimal transport problems $\mathcal{O}(N^3 log(N))$ \citep{Agueh2011}.

\subsubsection{Illustration of the application of barycenters to ensemble forecasts \label{ssec:MME_bary}}

The $L_2$ and $W_2$ distances have different properties in the distribution space, leading to distinct multi-model ensembles. 
The $L_2$ distance focuses on the ``vertical'' differences between distributions. This means that the $L_2$ distance between two distributions with disjoint supports is equal to the sum of their $L_2$ norms, no matter how distant their supports are (e.g., it is equal to $2\sum_i^N 1/N^2$ for two discrete distributions with each $N$ equiprobable points). As a result, the members of the $L_2$ barycenter always coincide with the members of the initial ensembles, regardless of the distance between the supports of the input distributions. This means that the members of the $L_2$ barycenter are physically consistent, according to the numerical weather model they were drawn from. 
In contrast, the $GaussW_2$ distance is a ``horizontal'' measure of the difference between two distributions, in the sense that it is based on the horizontal displacement (or mapping) between them \citep{Santambrogio2015}. This enables the associated $GaussW_2$ barycenter to have a different support than the input distributions. In practice, that means that the members of the $GaussW_2$ barycenter do not coincide with any of the input ensemble members. 
Although the covariance structures of the inputs are still taken into account, they are merged (step 2 of the method in Section~\ref{ssec:ot_bary}), allowing the barycenter to contain forecasts that could not have been generated by the individual models. The $GaussW_2$ barycenter is thus more flexible, but does not guarantee the same level of physical consistency as the $L_2$ barycenter.

\begin{figure}[ht]
  \noindent\includegraphics[width=\textwidth]{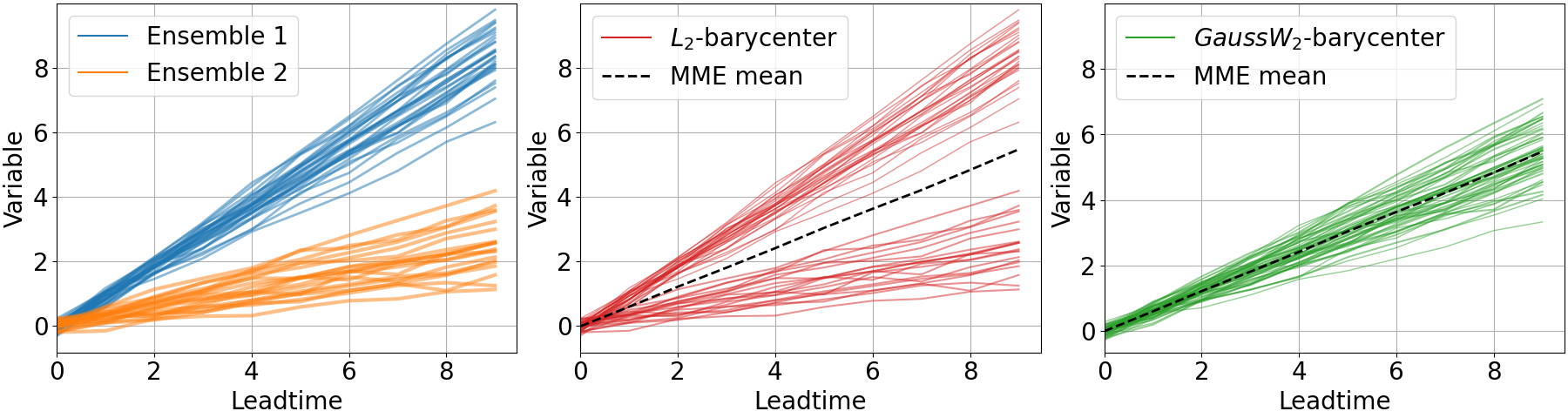}
  \caption{Illustration of (equal-weights) multi-model ensembles using barycenters for a synthetic variable. The ensemble can be seen as a discrete distribution whose points are the time series of each member. The weight of the points in the distribution is indicated here by the thickness of the line. }
  \label{fig:illustration_bary}
\end{figure}

Figure~\ref{fig:illustration_bary} gives an illustration of the two barycenters applied to the combination of two synthetic ensemble forecasts with $N_1=10$ and $N_2=15$ members respectively. 
The members of the $L_2$ barycenter in Figure~{\ref{fig:illustration_bary}b} correspond to the members of the input ensembles  in Figure~{\ref{fig:illustration_bary}a}.  The weight of each member of the $L_2$ barycenter is given by the weight of the corresponding member in the input ensemble ($1/N_1$ or $1/N_2$) multiplied by the corresponding model's weight (here $\lambda_1=\lambda_2=0.5$). 
The $GaussW_2$ barycenter shown in Fig.~{\ref{fig:illustration_bary}c} has a different structure compared to the $L_2$ barycenter. As expected, a major difference is that none of its members belong to the input ensembles. In other words, while the support of the $L_2$ barycenter is the union of the input supports, the support of the $GaussW_2$ barycenter lies along the path of the optimal transport and  can therefore be different. This also allows for the $GaussW_2$ barycenter to retain certain characteristics of the input distributions. For example, in this illustrative case, both input distributions are unimodal, and so is the $GaussW_2$ barycenter. This is not the case for the $L_2$ barycenter which has two modes, each associated with the mode of one of the input distributions. 
The two barycenters may be attractive for different types of forecasts. The $GaussW_2$ barycenter preserves some geometric properties of the input distributions, while the $L_2$ barycenter retains bimodality. 
Actual forecasts are more complex, and assessing the different barycenters will require extensive testing with different criteria. The purpose of Figure~\ref{fig:illustration_bary} is to emphasize the (hitherto largely untapped) variety of ways to build multi-model ensemble forecasts.

An important fact to note is that, despite their differences, the two barycenters have the same ensemble means (for the same model weights). Their mean is equal to the weighted mean of the input distributions's means (this follows easily from Eq.~\eqref{eq:l2_bary2} for the $L_2$ barycenter, and see \citet[Remark 9.1]{Peyre2018} for the $W_2$  and $GaussW_2$ barycenters). 
Thus, the difference between the barycenters is how they represent the forecast uncertainty. This raises the question: how different are the $L_2$ and $GaussW_2$ distributions, and which one captures forecast uncertainty better? 
We address this question empirically for a particular case study in the next sections. 
 

\section{Data and methodology}\label{sec:case_study}

In this study, we focus on subseasonal forecasts over Europe during the boreal winter. We consider one large-scale and two surface variables: the geopotential height at 500 hPa (z500), the daily temperature at 2 m (t2m), and the wind speed at 10 m (w10m). The latter two variables are interesting to predict for the energy sector, as energy demand is particularly dependent on the temperature in winter (due to the use of electrical heating), while the wind speed is an essential variable for estimating wind farm production. The geopotential height at 500 hPa is included as a traditional large-scale variable for comparison.

\subsection{Data}

To make predictions for this case study and to validate them, we need both a dataset of ensemble forecasts from multiple dynamical models, to which the barycenters will be applied and a reference dataset against which the forecast skill will be evaluated.

\subsubsection{S2S data \label{ssec:s2sdata}}

For this first implementation, we selected four models from the S2S database \citep{Vitart2017}: the European Centre for Medium-Range Weather Forecasts (ECMWF) model, the National Centers for Environmental Prediction (NCEP) model, the Environment and Climate Change Canada (ECCC) model, and the Korea Meteorological Administration (KMA) model (see Table~\ref{tab:model_description} for a description of their characteristics). An important criterion in model selection is their diversity. As previously explained, one advantage of using multi-model ensemble is to better sample model error. If the models have similar construction, this error is not sampled properly. Other criteria include time coverage and consistency in production over the study period.

\begin{figure}[!h]
\centering
\includegraphics[width=0.7\textwidth]{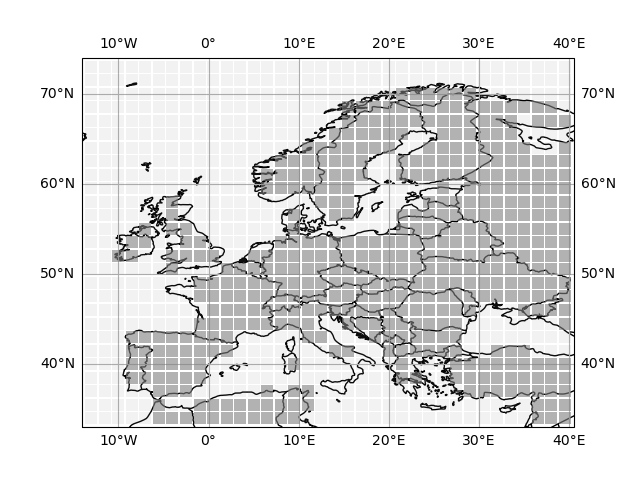}
\caption{Study domain with the $1.5^{\circ}\times1.5^{\circ}$ lat/lon grid from the forecasts. Only the land points (in grey) are used in this case study.}
\label{fig:domain}
\end{figure}

The forecasts and reforecasts from the selected models were retrieved from the S2S database through the ECMWF's Meteorological Archival and Retrieval System (MARS) directly on a $1.5^{\circ}\times 1.5^{\circ}$ grid for our study domain shown in Figure~\ref{fig:domain}, that is Europe ($34^{\circ}N-74^{\circ}N$, $13^\circ W$-$40^\circ E$). We retrieved the daily mean 2 m temperature, the instantaneous values of the 10 m wind speed, and the 500 hPa geopotential height at midnight (00Z). 
We selected the forecasts initialized during the winter months (December-January-February, DJF) for the 2016--2022 period, for which the four models have matching initialization dates. This resulted in a total of 90 simulations spanning seven winters. After pre-processing (described below), the forecasts were resampled to weekly means. The barycenter-based MME methods were applied to the weekly time series (i.e., $n_t=4$ weeks), which was also the temporal resolution used for forecast evaluation.

\paragraph*{Pre-processing:}
\begin{itemize} 
\item The forecasts from the four models are calibrated using the Mean and Variance Adjustment (MVA, \citet{Leung1999,Manzanas2019}) method, following \citet{Goutham2022}. Ensemble forecasts suffer from systematic errors (e.g. mean bias) due to uncertainties in ensemble initialization and model formulation. In particular, forecasts tend to drift away from the observed climate toward the  model climatology as lead time increases \citep{Takaya2019}, making statistical correction essential at extended-range time scales. 

The model climatology is estimated from reforecasts, which are ``retrospective'' forecasts generated with the same model but for past dates. The observed climatology is derived from a reference dataset (see Section~\ref{ssec:reference} below). The calibration method (MVA here) is then applied to statistically adjust the forecasts. To ensure a robust estimation of the climatologies, we aggregate all forecasts initialized within a 15-day window, comprising the same calendar day as the forecast starting date and the previous 14 days\footnote{previous 15 days for KMA to better match the frequency of its reforecasts} \citep{Manrique2020}. The observed climatology is computed on the same dates for consistency.

\item The NCEP and KMA models have smaller ensemble sizes but are produced daily. We thus create a lagged ensemble by aggregating the forecasts with those initialized the two preceding days for NCEP and the seven preceding days for KMA. This allows us to improve their skill and to increase their ensemble size, making them more comparable to the other models (48 and 32 members for NCEP and KMA,  respectively after lagging). 
The ECCC model has a smaller ensemble size than ECMWF and the lagged NCEP and KMA models. However, lagging is not applied to it because of its production frequency (it is only produced weekly).  
\end{itemize}

\begin{table}[ht]
\centering
\caption{Description of the four models from the S2S database (characteristics for the period used). See~\citet{Vitart2017} for more details. }\label{tab:model_description}
\begin{tabular}{p{2.6cm} p{3cm} p{3cm} p{3cm} p{3cm}}
\tophline
  & \multicolumn{1}{c}{ECMWF} & \multicolumn{1}{c}{NCEP}  & \multicolumn{1}{c}{ECCC}  & \multicolumn{1}{c}{KMA} \\ 
\middlehline
\textbf{Forecasts}    &            &               \\
~~~~Time range    &          d 0--46                 &           d 0--44  & d 0--32     & d 0--60       \\
~~~~Resolution    &           Tco639/Tco319 L137                &         T126 L64      & Yin-Yang grid at $35^\circ$ (varying vertical resolution) &  N216 L85          \\ 
~~~~Frequency     &   Twice a week  \newline (Monday, Thursday)   &  Daily   & Weekly \newline (Thursday) & Daily \\
~~~~Ensemble size \newline${}$~~~~(original) &              50+1             &            15+1      & 20+1 &   4     \\ 
~~~~Ensemble size \newline${}$~~~~after \newline${}$~~~~pre-processing &              51             &            48      & 21 &   32    \\ 
\middlehline
\textbf{Reforecasts}    &            &      &  &         \\
~~~~Method    &     On the fly       &       Fixed   & On the fly & On the fly     \\
~~~~Period   &     Past 20 years       &      1999--2010   & fixed period (varying) &   fixed period (varying)   \\
~~~~Frequency    &    Twice a week      &     Daily   & Weekly & 4/month       \\
~~~~Ensemble size &   10+1  &   3+1 & 3+1 & 3   \\
\bottomhline
\end{tabular}
\end{table}

\subsubsection{Reference data}\label{ssec:reference}
The Modern-Era Retrospective analysis for Research and Applications, version 2 (MERRA-2, \citet{Gelaro2017}) reanalysis is used as reference for calibration and forecast validation. We use a reanalysis as reference in order to have a spatially and temporally complete dataset. MERRA-2 was chosen because it is a recent reanalysis covering both the calibration and validation periods and is based on a different circulation model from those used in the selected S2S models. This would not be the case, for example, with the ERA5 reanalysis which is also produced by the ECMWF with a similar model as the S2S forecasts \citep{Herbach2020}. 
The three variables of interest were retrieved from NASA Goddard Earth Sciences (GES) Data and Information Services Center (DISC) on MERRA-2's native grid ($0.5^{\circ}$ lat $\times~0.625^{\circ}$ lon grid). The data were re-gridded to match the $1.5^{\circ}\times~1.5^{\circ}$ lat/lon grid of the forecasts using bilinear interpolation with the Climate Data Operators (CDO,~\citet{schulzweida_uwe_2022}). 

We also use MERRA-2 to build a 30-year rolling climatology. The climatology is a common benchmark for forecast validation, and it is also used to compute skill scores (see Section~\ref{ssec:scores}). For a given day of the year, the climatology consists of MERRA-2 values for the same day over the previous 30 years. 
The climatology is thus an ensemble with 30 members, each corresponding to a different year.


\subsection{Evaluation} \label{ssec:scores} 

We now introduce the evaluation scores used in this study. 
We use the Continuous Ranked Probability Score and related skill scores to evaluate the accuracy of the forecasts, and the Spread-Skill Ratio to investigate their calibration. Due to the low number of samples in our case study (90 simulations), we do not investigate the prediction of extreme events. 
All scores used in this study are summarized in Table~\ref{tab:scores}.

\subsubsection{Continuous Ranked Probability Score (CRPS)}
The {CRPS} is a widely used score for probabilistic forecasts of continuous variables \citep{Matheson1976,Gneiting2007,Wilks2019_chap9}. The CRPS is actually the squared $L_2$ distance between the Cumulative Distribution Function (CDF) of the ensemble forecast, and the CDF of the observation: 
\begin{align}
     \text{CRPS}_n & =   \int_{-\infty}^{+\infty}\left[F_{\text{fc},n}(y) - F_{\text{ref},n}(y) \right]^2 dy \label{eq:CRPS}
\end{align}
where $F_\text{fc}$ is the CDF of the forecast, $F_\text{ref}$ is the CDF of the observation and $n\in[\![1, N]\!]$ is the ensemble member index. The CDFs are computed empirically from the ensembles. In general the reference is deterministic, and its CDF is then a step function 
$F_\text{ref}(y) = \mathbf{1}_{y \geq x_{\text{ref}}}$, 
where $x_{\text{ref}}$ is the deterministic value of the reference. 

The accuracy of the ensembles is evaluated by computing the mean CRPS (over the forecasts indexed by~$n$). For easier comparison across variables, we use  the Continuous Ranked Probability Skill Score (CRPSS) which compares the mean CRPS of the ensemble forecast to the mean CRPS of the climatology, but statistical tests will be performed on the CRPS distribution. 

We consider two other scores based on the CRPS.  
To evaluate robustness to outliers, we use the proportion of skillful forecasts (hereafter CRPSp), which corresponds to the percentage of simulations for which the model has a better CRPS than the climatology \citep{Goutham2022}. We consider that the model has some skill if $50\%$ of the forecasts are skillful.  
As a counterpart, we introduce a third score that focuses on the tail of the distribution (i.e. forecasts far from the truth). The proportion of critical failures (CRPSf) is defined as the percentage of simulations for which the model has a $\text{CRPS}_n$ larger than twice that of the climatology (this metric is thus negatively oriented, contrary to the CRPSS and CRPSp). The CRPSp and CRPSf should be interpreted together.

\bigskip
\noindent\textbf{Remarks} on the links between the CRPS and the $L_2$ barycenter:
\begin{itemize}
    \item It is interesting to note that for univariate distributions, the $L_2$ barycenter is also the barycenter with respect to the energy distance (see appendix~B). Thus, since the CRPS is identical to the energy distance in 1D \citep[see][]{Wilks2019_chap9}, we can expect good CRPS performance for the $L_2$ barycenter. 

    For symmetry, the Wasserstein distance could also be used as a score for forecast validation. However, it is equivalent to the RMSE (averaged over the members) in the case of a deterministic observation and so it does not evaluate well the representation of the uncertainties in the forecasts. Thus, we do not use it here.
    
    \item The CRPS of the $L_2$ barycenter $\mu_{L_2}^{\alpha}$ of the distributions $(\mu_1,\ldots,\mu_d)$ (with weights $\lambda_1,\ldots,\lambda_d$) can be expressed as a function of their CRPS with respect to the observation and their cross-CRPS:
    \begin{align}
        \text{CRPS}(\mu_{L2},\nu_{obs}) & = \sum_{i=1}^{n}\lambda_i \text{CRPS}(\mu_i,\nu_{obs}) - \sum_{i=1}^{n} \sum_{j\neq i} \frac{1}{2}\lambda_i \lambda_j \text{CRPS}(\mu_i,\mu_j)  \label{eq:crps_l2}
    \end{align}
where $\nu_{obs}$ is the observation (see appendix~C). This expression clearly shows that the CRPS of the $L_2$ barycenter is always smaller than the average CRPS of the input distributions. In fact, the more different the input distributions are (i.e., large cross-CRPS), the more advantageous it is to merge them.
\end{itemize}

\subsubsection{Spread-skill ratio}
The CRPS evaluates the accuracy of the forecasts, that is, their overall quality \citep{Wilks2019_chap9}. In order to focus on the calibration (or reliability) of the ensembles, we also look at the spread-skill ratio (SSR) defined as the ratio of the ensemble standard deviation to the root mean square error RMSE of the ensemble mean. 
A perfectly reliable ensemble has a SSR equal to one, while larger SSR values indicates overdispersion and smaller values underdispersion \citep{Fortin2014}. The SSR can be used as a first evaluation of an ensemble's reliability (a value of 1 is necessary, though not sufficient, for a fully reliable ensemble).

\begin{table}[!ht]
\center
\caption{Scores used for evaluation.}\label{tab:scores}
\begin{tabular}{llll}
\tophline
Name & Formula & Range & Comment   \\
\middlehline
\middlehline
$\overline{\text{CRPSm}}_l$ & \(\displaystyle \frac{1}{N_{for}} \sum_{k=1}^{N_{for}} \langle  \text{CRPS}_{\text{fc},k,l} \rangle \) & $\left[0,+\infty \right)$  & negatively oriented   \\
CRPSS & \(\displaystyle 1 - \frac{\overline{\text{CRPSm}}_l(\text{fc})}{\overline{\text{CRPSm}}_l(\text{clim})} \) & $\left(-\infty, 1 \right]$ & positively oriented   \\
CRPSp & \(\displaystyle \frac{1}{N_{for}} \sum_{k=1}^{N_{for}}  \frac{100}{N_{lat} N_{lon}} \sum_{i=1}^{N_{lat}} \sum_{j=1}^{N_{lon}}  \left[ \text{CRPS}_{\text{fc},k,i,j,l} < \text{CRPS}_{\text{clim},k,i,j,l} \right] \) & $\left[0, 100 \right]$ & positively oriented   \\
CRPSf & \(\displaystyle \frac{1}{N_{for}} \sum_{k=1}^{N_{for}}  \frac{1}{N_{lat} N_{lon}} \sum_{i=1}^{N_{lat}} \sum_{j=1}^{N_{lon}}  \left[ \text{CRPS}_{\text{fc},k,i,j,l} > 2~ \text{CRPS}_{\text{clim},k,i,j,l} \right] \) & $\left[0, 100 \right]$  & negatively oriented   \\
\middlehline
SSR & \(\displaystyle
    SSR = \frac{\sqrt{\frac{1}{N_{for}} \sum_{k=1}^N \langle \text{var}_m (\text{fc}_{k,m})\rangle }}{\sqrt{\frac{1}{N_{for} } \sum_{k=1}^N \langle \left( \text{mean}_m(\text{fc}_{k,m}) - \text{obs}_{k} \right)^2 \rangle }} \) & $\left[0,+\infty \right)$ & \makecell{SSR$>1$: overdispersion \\ SSR$<1$: underdipersion}   \\ 
\middlehline
Notations & \multicolumn{3}{l}{\makecell{ $i\in \llbracket 1,N_{lat}\rrbracket$: latitude; $j\in \llbracket 1,N_{lon}\rrbracket$: longitude; $l\in \llbracket 1,N_{day}\rrbracket$: lead time; \\ $k\in \llbracket 1,N_{for}\rrbracket$: forecast (i.e. starting date); $m\in \llbracket 1,N_{mem}\rrbracket$: member. }}   \\
Remarks & \multicolumn{3}{p{12cm}}{ - $\langle.\rangle$ denote the area-weighted spatial average. \newline - $\text{mean}_m$ and $\text{var}_m$ are the mean and the variance over the ensemble members respectively.}   \\
\bottomhline
\end{tabular}
\end{table}


\section{Results \label{sec:results}}

In this section, we evaluate the combination of four models' ensembles applying the two barycenter-based MME methods described in Section~\ref{sec:MME} to the case study described in Section~\ref{sec:case_study}. All the results shown here are obtained for barycenters with equal weights (that is $\lambda_\text{ECMWF},\lambda_\text{NCEP},\lambda_\text{ECCC},\lambda_\text{KMA}=1/4$). We tested optimizing the weights with respect to the CRPS, but it leads to either little improvement or slight deterioration of the forecast skill depending on the variables and scores (not shown here).

\subsection{Validation and comparison of the barycenter-based MME \label{ssec:result1}}

\begin{figure}[!ht]
\centering
\includegraphics[width=\textwidth]{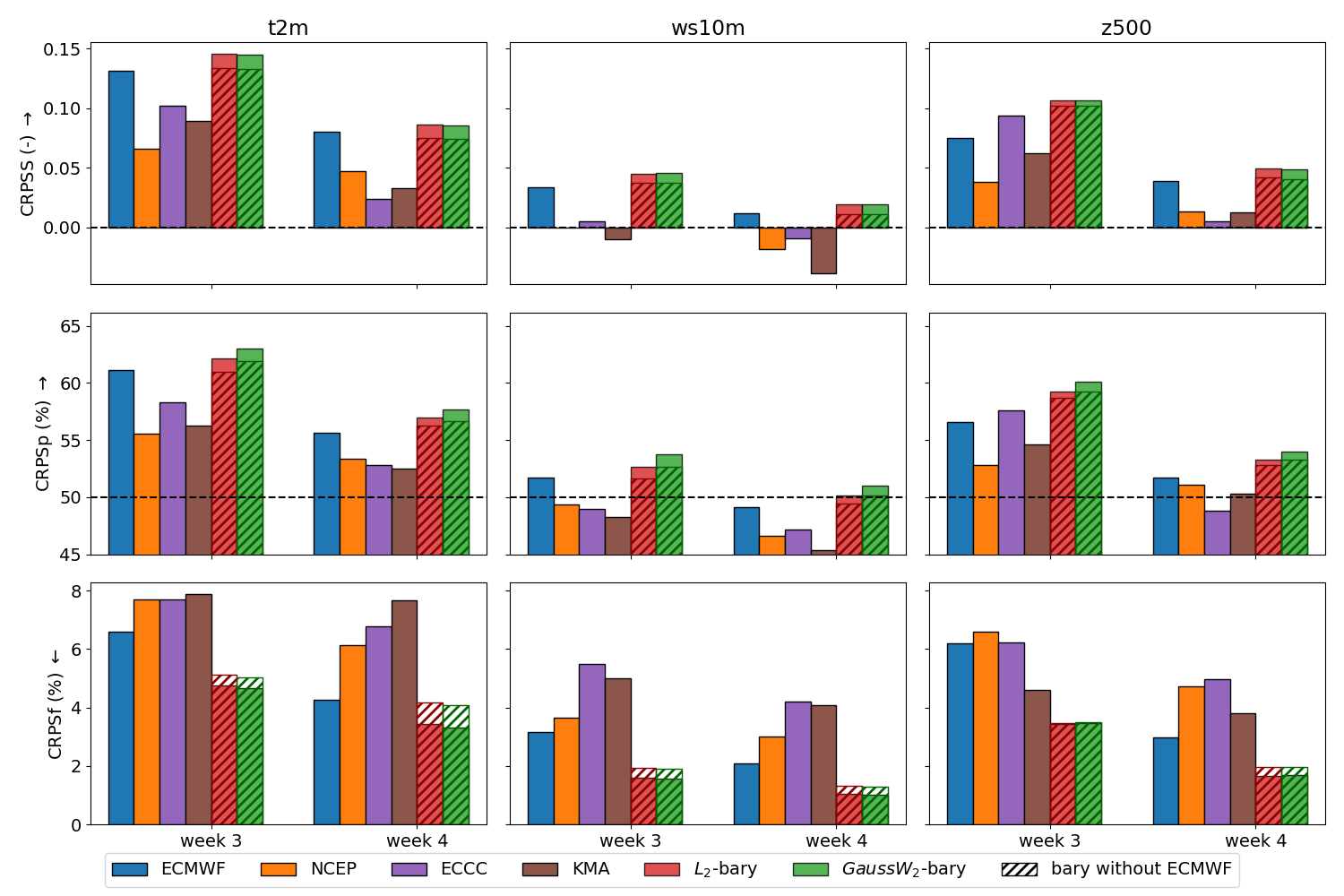}
\caption{Skill scores for the weekly 2m temperature (left), 10m wind speed (center) and geopotential height at 500hPa (right): the CRPSS (top), the proportion of skillful forecasts (CRPSp, middle) and proportion of critical failure (CRPSf, bottom).}
\label{fig:summary_score}
\end{figure}

Figure~\ref{fig:summary_score} summarizes the performance of the four SMEs and the two barycenter-based MMEs. For the latter, the bars represent the performance of the barycenters combining the four SMEs (the hatched areas represent the performance of the barycenters of three SMEs out of four; these results are discussed below, Sect.~\ref{ssec:withoutECMWF}). 
The distribution of the scores (CRPSS, CRPSp and CRPSf) over the starting dates is represented by its mean in the plot, and is used to test if the models' scores are significantly different from each other at a $5\%$ significance level. The significance is estimated using the Wilcoxon signed-rank test, a non-parametric paired statistical test \citep{Wilcoxon1945}. We chose to show the CRPSS here instead of the mean CRPS for an easier comparison across variables. However, the mean and the statistical tests are computed on the (spatially averaged) CRPS distributions. The p-values of these statistical tests are shown in Appendix~\ref{app:wilcoxon}.

For all three variables and three scores, the two barycenters outperformed the SMEs, however the gain with respect to the best performing SME is sometimes small. In particular, ECMWF has the best performance among the four SMEs for the surface variables, the 2 m temperature and the 10 m wind speed. In terms of CRPSS, the two barycenters have similar values and are significantly different from all the SMEs for the 10 m wind speed and the geopotential height (see Tables~\ref{tab:sign_w10m} and~\ref{tab:sign_z500}). Even though two or three of the models (depending on the week) have a negative CRPSS for the 10 m wind speed, merging them with more skillful SMEs still adds value: the barycenters have a significantly better CRPSS. 
For the 2 m temperature, the barycenters do not have a significantly different CRPSS than ECMWF, which already has a relatively large CRPSS (see Table~\ref{tab:sign_t2m}). 

In terms of CRPSp, the $GaussW_2$ barycenter is significantly better than the others, including the  $L_2$ barycenter that ranks second (Tables~\ref{tab:sign_t2m}, \ref{tab:sign_w10m} and ~\ref{tab:sign_z500}). That is, the barycenters, and especially the $GaussW_2$ barycenter, are more often better than the climatology compared to the SMEs. 
The barycenter-based ensembles also significantly reduce the risk of large failures as shown by their lower CRPSf. The two barycenters have similar CRPSf values, clearly lower than the values of the SMEs. 
In particular, the proportion of critical failures is divided by almost two for the 10 m wind speed compared to the best SME, ECMWF.

\begin{figure}[!ht]
\centering
\includegraphics[width=\textwidth]{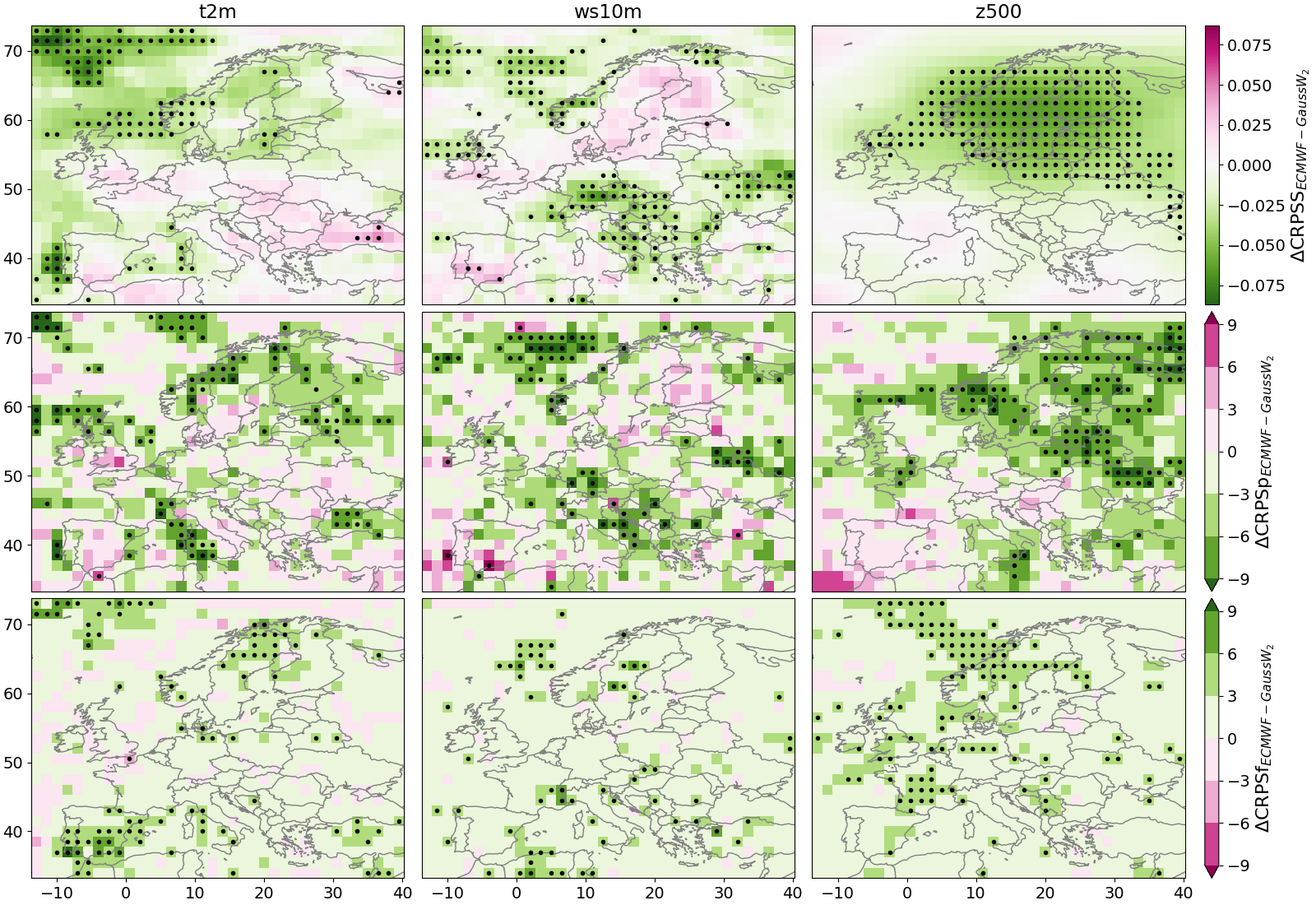}
\caption{Maps of the differences between ECMWF and $GaussW_2$ barycenter in terms of (top) CRPSS, (middle) CRPSp, and (bottom) CRPSf averaged over weeks~3 and~4. Green indicates that $GaussW_2$ barycenter performs better and pink that ECMWF does. For readability reason, a discrete colorbar is used for the $\Delta$CRPSp and $\Delta$CRPSf which are more pixelized. Dots indicate grid points where the difference is statistically significant at the $5\%$ level (using a two-sided Wilcoxon signed rank test for the CRPS and a McNemar test for the CRPSp and CRPSf).}
\label{fig:spatial_scores}
\end{figure}

Variations in scores across the spatial domain for each model are large, in the sense that they tend to dominate over differences between models (not shown here). We thus expect this to be also the case for the MMEs.
Figure~\ref{fig:spatial_scores} shows the differences between ECMWF and the $GaussW_2$ barycenter's scores. 
We chose ECMWF as a reference, since it has the best performance among the SMEs for two of the three variables (the surface ones). The map of the differences between the $L_2$ barycenter and ECMWF is very similar, and so is not shown here. 
For the CRPSS, one can see clear areas that benefit from the MMEs, but these areas are not the same depending on the variables. For the 2 m temperature, the barycenter is more advantageous in the North-West, above the North Sea, Atlantic Ocean and Scandinavia.
ECMWF is better above Sweden and Finland for the 10 m wind speed.
For this variable, the barycenter benefits more in central Europe as well as southern and eastern Europe.
The spatial pattern is smoother for the geopotential height, with the largest differences above northern Europe, where the barycenter is more skillful. 
Despite these spatial variations across the three variables, the barycenter outperforms ECMWF at a majority of grid points.

The maps for the CRPSp differences are more pixelated, however the positive and negative areas roughly match those of the CRPSS differences. 
In contrast, the maps of the CRPSf differences show very different spatial patterns.
For example, ECMWF performs better over Spain for the two surface variables for both the CRPSS and the CRPSp, but the barycenter has a better CRPSf there. In fact, the barycenter outperforms ECMWF in terms of CRPSf almost everywhere. 
This means that combining ensemble forecasts has a stabilizing effect, avoiding extremely bad forecasts. 
However, the differences are relatively small, less than  $3\%$ at most grid points.

\subsection{Ensembles calibration}

\begin{figure}[!ht]
\centering
\includegraphics[height=0.85\textheight]{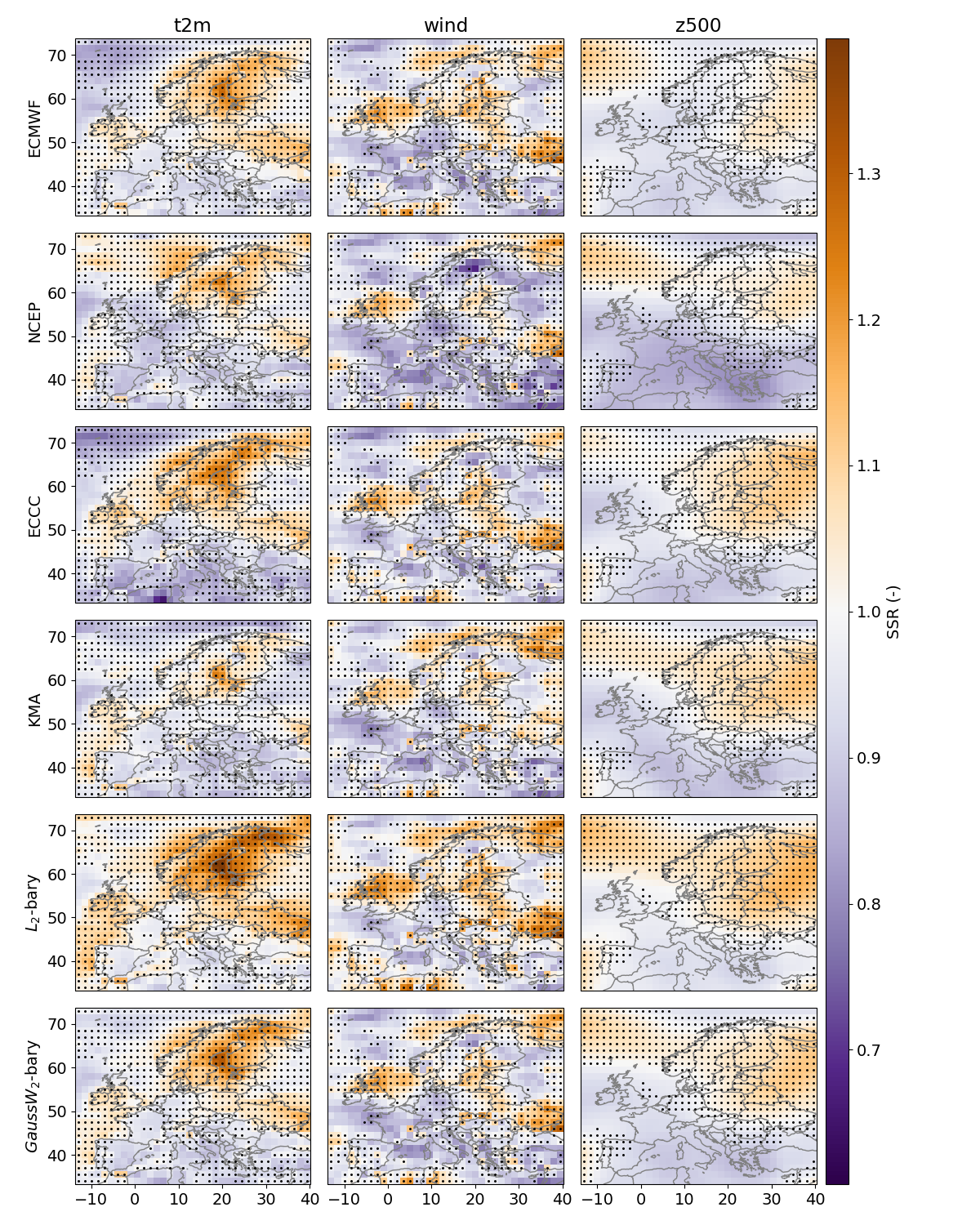}
\caption{Spread-Skill-Ratio (SSR) for the weekly 2m temperature (left), 10m wind speed (center) and geopotential height at 500hPa (right) averaged over weeks~3 and~4 for the four single-model ensembles and the two barycenters. Orange indicates overdispersion (i.e. under-confident forecasts), and blue under-dispersion (i.e. overconfident forecasts). Dots indicate grid points where the ensemble variance and MSE of the ensemble mean are are statistically  different  at the $5\%$ level (using a two-sided Wilcoxon signed rank test on the distribution of variance and MSE).}
\label{fig:spatial_SSR}
\end{figure}

In order to get insight into the calibration of the ensemble forecasts, we look at their SSR maps in Figure~\ref{fig:spatial_SSR}. The spatial patterns of weeks~3 and~4 being similar, only their average is shown here. The spatial patterns are consistent across the SMEs. They tend to be under- or over-confident in the same regions, but with different intensities. For example, all models are over-dispersed over Scandinavia for the 2m temperature forecasts, with KMA having the least and ECCC the most overdispersion. On the contrary, they are all under-dispersed over the Mediterranean Sea. 
 
Note that, by construction, the barycenters have the same ensemble mean and so the same RMSE, so that the differences in their SSR are only due to their spread. Their SSRs have similar patterns as the SMEs. However, one can notice that the $L_2$ barycenter tends to have more spread than the $GaussW_2$ barycenter: it shows more overdispersion and less underdispersion. This can be explained by the way they were constructed.
The variance of the $L_2$ barycenter is indeed given by
\begin{equation}\label{eq:var_l2}
\text{Var}\left(\mu_{L_2}\right) = \sum_{i=1}^{d} \lambda_i\text{Var}\left(\mu_i\right) + \sum_{i=1}^k  \lambda_i \left(\overline{\mu_i}- \sum_{j=1}^k \lambda_j \overline{\mu_j}\right)^2,
\end{equation}
where $\overline{\mu}_i$ is the mean of the distribution $\mu_i$.  
The variance of the barycenter is thus always equal to or higher than the average of the SME's variances. This property is advantageous in the case of under-dispersed input ensembles. However, it is a disadvantage if they are better calibrated or already over-dispersed \citep[e.g.][]{Zhou2022,Eade2014}. 
In those cases, the $GaussW_2$ barycenter is more advantageous. 
In order to build it, the input ensembles are first centered and their covariances scaled before being pooled together. Thus, its covariance is the one of the Gaussian barycenter (Sect.~\ref{ssec:ot_bary}).

\FloatBarrier
\subsection{MME without ECMWF}
\label{ssec:withoutECMWF}

In Section~\ref{ssec:result1}, ECMWF is found to be more skillful than the other three SMEs for the two surface variables. Moreover, although the two barycenters have overall better performance, ECMWF is still better in some regions. Thus, to investigate the importance of ECMWF within the combination, we compute the barycenters of NCEP, ECCC and KMA (without ECMWF). They are compared to the barycenters with ECMWF in Figure~\ref{fig:summary_score} (hatched bars).
As expected, we observe a decrease in performance for all scores and variables.
This is also true for most grid points when looking at the maps of differences except for those points where the barycenters (with four models) are notably more skillful than ECMWF (not shown here).

When comparing the barycenters with the three SMEs used to build them, we can make similar observations as in Section~\ref{ssec:result1}. Specifically, the $L_2$ and $GaussW_2$ barycenters have similar CRPSS and CRPSf, but the $GaussW_2$ barycenter has a significantly better CRPSp (at the $5\%$ level, Tables~\ref{tab:sign_t2m_3m}, \ref{tab:sign_w10m_3m} and \ref{tab:sign_z500_3m}). Both barycenters also outperform the three SMEs for all metrics. 
However, two features are noteworthy. First, the CRPSS and CRPSp differences between the barycenters and the SMEs for the surface variables are notably larger than those observed between the barycenters with four models and ECMWF.
A particularly striking example is the CRPSS of the 10 m wind speed: while the three SMEs show little to no skill, with very low or even negative CRPSS, their barycenters have CRPSS similar to ECMWF's.
Second, the barycenters with only three models are performing as well as ECMWF in terms of CRPSS and CRPSp (there are no significant statistical differences, see Appendix~\ref{app:sign_3m}) and better in terms of CRPSf. Thus, with three less skillful models, one can reproduce the considerably better performance of ECMWF in the case studied here.

\FloatBarrier

\section{Discussion\label{sec:discussion}}

\subsection{$L_2$ barycenter versus $GaussW_ 2$ barycenter}

We have seen that the MMEs improve the average forecast skill compared to the SMEs with respect to all scores and for all the variables considered, even though a SME may perform better at some locations.
In general, the two barycenters have similar CRPSS and CRPSf, but the $GaussW_2$ barycenter has a better CRPSp.
In other words, the two barycenters are similar on average, but the $GaussW_2$ is better more often  without leading to an increase in very poor forecasts (with respect to the CRPS). 
Another important difference between the barycenters is the way they represent forecast uncertainty through their ensemble spreads. The $L_2$ barycenter always has a larger ensemble spread than the $GaussW_2$ barycenter, which in turn has an impact on their reliability.

The two barycenters have the same means (which only depend on the means of the input distributions), but their covariances differ.
The covariances of the Wasserstein barycenter depend solely on the covariances of the input ensembles (see Section~\ref{sec:MME}), while the covariances of the $L_2$ barycenter depend on both their means and their variances. 
If we assume that the variances of the input ensembles represent the uncertainty due to their initial conditions, their average in the formula of the $L_2$ barycenter variance (first term of Eq.~\eqref{eq:var_l2})  can be interpreted as a measure of the MME's initial-condition uncertainty. The second term, the variance of the input ensemble means, can be seen as a measure of model uncertainty as it quantifies the differences between ensemble (means).
Thus, the variance of the $L_2$ barycenter accounts for model errors but ignores differences in higher moments than the ensemble mean. 
On the other hand, the $GaussW_2$ barycenter merges the forecast uncertainties of the different models (i.e. the uncertainty due to the initial conditions).
Instead, information about model uncertainty lies in the value of the barycenter distance: $\min_{\mu}\sum_{k=1}^{d}\lambda_k.d(\mu_k,\mu)^2$. 
This information is not exploited here, but it could be considered to account for model uncertainty.
Finally, not only are dynamical and model uncertainties captured by different quantities for the two barycenters (variance of the barycenter versus barycenter distance), but the way they are measured also differs.
Understanding the implications of these differences requires further investigation.

\subsection{Best model versus models combination}

There can be several reasons why merging ECMWF with less skillful models leads to barycenters with improved performance. 
First, the three SMEs have generally less skill than ECMWF, but can be occasionally better for some given locations, lead times, or initialization dates. The barycenters can exploit this information to improve skill thanks to error cancellation and to the non-linearity of the skill metrics \citep{Hagedorn2005}.
Second, the ECMWF forecasts may have good performance but be overconfident. In that case, adding other models with lower skill increases the spread of the ensemble and can move the ensemble mean towards the truth, as shown by~\citet{Weigel2008} for seasonal ensemble forecasts. 
For the $L_2$ barycenter, this can be seen from Equation~\eqref{eq:crps_l2} which shows that the CRPS of the $L_2$ barycenter is composed of two parts: the weighted average of the CRPS of the SMEs minus the weighted average of the CRPS between all pairs of SMEs. Thus, if one model has a worse CRPS than another, it can still improve the CRPS of the barycenter if the CRPS between the models is large enough to compensate for its own CRPS.

\subsection{Weighted barycenters}

Equal weighting of the models in the pooling method is the simplest and most used approach.
However, some studies investigate the use of weighted multi-model ensembles with mixed results (at the weather or seasonal scale in~\citet{Weigel2008,Casanova2009,Kharin2002} and at the climate scale in~\citet{Haughton2015}). At the subseasonal scale,~\citet{Wanders2016} used mutivariable linear regression on the ensemble means to derive the multi-model weights. They show that the weighted muti-model ensemble has better deterministic performance but also better probabilistic performance (in terms of the Brier score) than the unweighted multi-model ensemble. 

Here, we have shown results using equal weights for the models, but the barycenter formulation easily allows for the construction of weighted MMEs. 
Results for weights that do not depend on space or time, minimizing the mean CRPS, were also obtained but not shown because, contrary to \citet{Wanders2016}, no significant improvement was found.
One explanation could be the limited time period used here. The collection of forecasts by the S2S database started in 2015. We thus have less than ten years of data, which may not be enough to find stable weights, as shown by~\citet{Wanders2016, Kharin2002}. 
Using reforecasts (instead of forecasts) would allow one to extend the time period.
However, it would still be limited to twelve years (due to the limited NCEP reforecast period) and would lead to additional difficulties.
In particular, the starting dates of the reforecasts for the different models in the S2S database do not match. In addition, the reforecasts have fewer members than the forecasts, so the reforecast results may not be transferred to the forecasts without additional assumptions.

Yet, another strategy could be to allow the weights to depend on space.
We indeed saw that even if the barycenters tend to be more skillful than the SMEs, they can be outperformed by ECMWF for some scores at some locations.
This suggests that giving more weight to ECMWF in these regions would be beneficial. Similarly, even if the barycenter without ECMWF showed a general degradation of the scores compared to the barycenters with the four models, it did perform better in some specific regions.
Thus, the barycenter with four models would benefit from giving less or no weight to ECMWF in these regions.
However, optimizing weights grid point by grid point requires more data and may lead to spatial inconsistencies~\citep{Wanders2016,DelSole2013}.
An alternative would be to learn weights per region (such as in \citet{Wanders2016}) or to enforce some spatial smoothness on the weights (also acting as regularization and helping with overfitting). 


\section{Conclusion}
\label{sec:conclusion}

We explore methods to combine ensemble forecasts from multiple models based on barycenters of forecast ensembles. Building on the recognition of the relevance of probabilistic forecasts for S2S prediction, we work directly in the probability distribution space. 
That is, the ensemble forecasts are manipulated as discrete probability distributions. This allows us to use existing tools from this space and, in particular, the notion of barycenters. 
Here, we explore two barycenters based on different metrics: the $L_2$ distance and the Wasserstein ($GaussW_2$) distance. 
We show that the $L_2$ barycenter is in fact equivalent to the well-known pooling method and compare it to the new $GaussW_2$ barycenter-based method.
Moreover, the variance of the $L_2$ barycenter can be decomposed into a term representing the average uncertainty of individual forecasts and a term representing model systematic error, while the variance of the $GaussW_2$ barycenter does not account for the latter. Instead, this information is captured by the values of the $GaussW_2$ barycenter. 

This first application of our framework to S2S prediction is illustrated through the combination of four single-model ensembles to predict winter surface temperature, surface wind speed, and 500 hPa geopotential height over Europe.
By construction, the two barycenters have the same ensemble means, which means they have the same performance in terms of deterministic scores. 
However, they differ in the way they represent forecast uncertainty. In particular, by construction, the $L_2$ barycenter is likely to be underconfident due to overdispersion, implying that the $L_2$ barycenter is advantageous in the case of under-dispersive input ensembles, but disadvantageous in terms of calibration otherwise.
More generally, we show that both barycenters perform similarly on average but that the $GaussW_2$ barycenter performs better more often (across the different grid points and dates) with respect to the CRPS.
Even in cases where one of the SMEs has superior skill compared to the others (ECMWF for the surface variables), it is still advantageous to combine them into a barycenter. This confirms the interest of multi-model methods shown by previous studies. We also show that by using the other three models, one can build barycenter-based MMEs that are as skillful as the best SME. 

This study is a proof of concept to develop the framework and investigate the properties of the barycenter-based MMEs. 
These results constitute a promising first step towards improving S2S predictions using barycenters to merge ensemble forecasts. A next step would be to further investigate the optimization of the weights in the barycenter in order to build weighted MMEs. Another interesting question concerns the possibility of building multivariate Wasserstein barycenters. The $L_2$ barycenter does not take into account the covariances between the variables, whereas the Wasserstein barycenter does. Finally, the implications of the fact that both barycenters account differently for forecast and model uncertainty deserve further investigation.


\section*{Code and data availability}
The Python code for the barycenter-based MMEs is openly available on Github (\url{https://github.com/clecoz/OT_for_MME}, last access: 20/03/2025) and Zenodo (\url{https://zenodo.org/records/15058503}, \citet{code_zenodo}, last access: 17/09/2024). 
\newline
This work is based on S2S data. S2S is a joint initiative of the World Weather Research Programme (WWRP) and the World Climate Research Programme (WCRP). The original S2S database is hosted at ECMWF as an extension of the TIGGE database (\url{https://apps.ecmwf.int/datasets/data/s2s/}). MERRA-2 data is available from the Goddard Earth Sciences Data and Information Services Center (GES DISC) data archive \citep[see][]{Merra2_t2m, Merra2_wind, Merra2_z500}. 
The pre-processed data are shared on Zenodo (\url{https://doi.org/10.5281/zenodo.15038871}, \citet{data_zenodo}).

\section*{Acknowledgment}
The authors thank Naveen Goutham for code and advice on forecast calibration. 
The authors thank the institut de Mathématiques pour la Planète Terre for (partial) funding (iMPT-AAP2021). 
This work was partially supported by the grant ANR-23-ERCC-0006 from Agence nationale de la recherche (ANR). 
This work is supported by Hi! PARIS and ANR/France 2030 program (ANR-23-IACL-0005). 
This research was produced within the framework of Energy4Climate Interdisciplinary Center (E4C) of IP Paris and Ecole des Ponts ParisTech. This research was supported by 3rd Programme d’Investissements d’Avenir (ANR-18-EUR-0006-02).


\newpage

\bibliography{references} 

\appendix

\section{Variance shrinkage in discrete Wasserstein barycenter}
\label{app:variance_shrinkage}

\begin{figure}[!ht]
\centering
\includegraphics[width=0.5\textwidth]{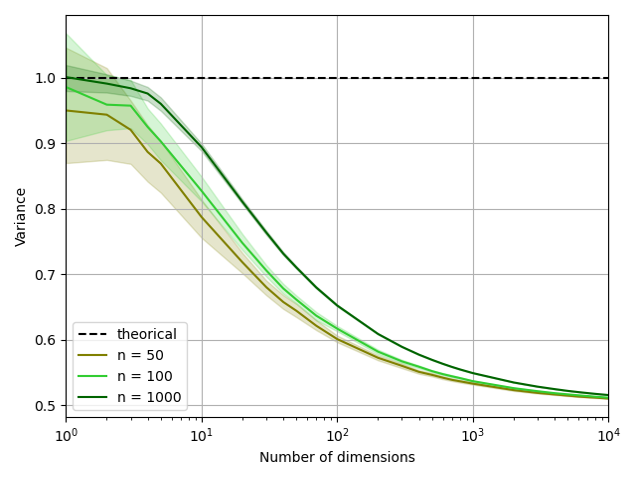}
\caption{Average variance of the discrete Wasserstein barycenter as a function of the dimension $d$ for different sample size $n$.  The full line represents this variance averaged over the 100 iterations with the minimum and maximum values indicated by the envelop.
}
\label{fig:var_shrinkage}
\end{figure}

A potential problem of the discrete Wasserstein barycenter is its reduced variance. Indeed, computing the Wasserstein barycenter directly on discrete probability distributions leads to a variance shrinkage problem which worsens with increasing space dimension and decreasing distribution sampling. This problem can be illustrated with the following empirical experiment:
\begin{enumerate}
    \item Create two discrete distributions by randomly drawing $n$ samples from two normal distributions with unit-covariance in $\mathbb{R}^d$.
    \item Compute the Wasserstein barycenter of the two discrete distributions (using the closed-form expression).
    \item Compute the normalized trace of the covariance of the barycenter divided by the number of variables (that is, the average variance of the variables).
    \item Repeat the three precedent steps 100 times.
\end{enumerate}
We repeated this experiment for different sampling sizes $n$ and number of dimensions $d$. The results of this experiment are shown in Figure~\ref{fig:var_shrinkage}. 
By construction, the Wasserstein barycenter of the two normal distributions with unit covariance also has a unit-variance (and so an average variance of one). However, one can observe that the discrete barycenter has a lower average variance. The average variance shrinks as the sample size decreases and the number of dimensions increases. For four dimensions and 50 samples, we observe a loss of more than $10\%$ of the average variance. This setting is comparable to our application, where the number of dimensions $d$ corresponds to the number of lead times (i.e. four weeks), and the number of samples $n$ corresponds to the ensemble size (ranging from 21 to 51 members, see Section~\ref{ssec:s2sdata}).

\FloatBarrier
 
\section{Energy distance and its associated barycenter \label{app:energy_dist}}

Let $\mu_1$ and $\mu_2$ be two distributions, and $F_1$ and $F_2$ be their CDF.\@ That is,  $F_1(x) = \int_{-\infty}^{x}\mu_1(t) dt$ and $F_2(x) = \int_{-\infty}^{x}\mu_2(t) dt$ $\forall x\in\mathbb{R}$.
The squared energy distance between $\mu_1$ and $\mu_2$ is 
\begin{align}
    \mathcal{E}^2(\mu_1,\mu_2) & = \int_{-\infty}^{+\infty}\left[F_1(x) - F_2(x) \right]^2 dx
      = \int_{-\infty}^{+\infty}\left[\int_{-\infty}^{x}\mu_1(t) dt - \int_{-\infty}^{x}\mu_2(t) dt \right]^2 dx \nonumber
\end{align}

The energy barycenter of $\mu_1$ and $\mu_2$ is the solution of the following minimization problem $ \mu_\mathcal{E}^\alpha = \argmin_{\mu} \alpha \mathcal{E}^2(\mu,\mu_1) + (1-\alpha) \mathcal{E}^2(\mu,\mu_2)$. Let ${\mathcal{B}(\mu)}=\alpha \mathcal{E}^2(\mu,\mu_1) + (1-\alpha) \mathcal{E}^2(\mu,\mu_2)$, then 
we have:
\begin{align*}
    \frac{d}{d\mu}\mathcal{B}(\mu) & = \frac{d}{d\mu} \left[  \alpha \int_{-\infty}^{+\infty}\left[F(x) - F_1(x) \right]^2 dx + (1-\alpha) \int_{-\infty}^{+\infty}\left[F(x) - F_2(x) \right]^2 dx  \right] \\
    &= \frac{d}{d\mu}  \int_{-\infty}^{+\infty}\left( \alpha\left[\int_{-\infty}^{x}\mu(t) dt - \int_{-\infty}^{x}\mu_1(t) dt \right]^2 + (1-\alpha) \left[ \int_{-\infty}^{x}\mu(t) dt - \int_{-\infty}^{x}\mu_2(t) dt \right]^2  \right) dx \\
    & = \int_{-\infty}^{+\infty}\frac{d}{d\mu}  \left( \alpha\left[\int_{-\infty}^{x}\left(\mu(t) - \mu_1(t)\right) dt \right]^2 + (1-\alpha) \left[ \int_{-\infty}^{x}\left(\mu(t) - \mu_2(t)\right) dt \right]^2  \right) dx \\
    & = \int_{-\infty}^{+\infty} \left(2\alpha \int_{-\infty}^{x} \left[\mu(t) - \mu_1(t)\right]dt \mu(x)  \right)dx +  \int_{-\infty}^{+\infty} \left(2 (1-\alpha) \int_{-\infty}^{x} \left[\mu(t) - \mu_2(t)\right]dt \mu(x)  \right)dx \\
    & = 2 \int_{-\infty}^{+\infty} \left(  \int_{-\infty}^{x}  \left(\mu(t) - \left[\alpha \mu_1(t) + (1-\alpha) \mu_2(t)\right]\right)dt\right) \mu(x) dx
\end{align*}

Thus, 
\begin{align*}
    \frac{d}{d\mu}\mathcal{B}(\mu_{L_2}^\alpha) & =0
\end{align*}
where $\mu_{L_2}^\alpha = \alpha \mu_1 + (1-\alpha) \mu_2$ is the $L_2$ barycenter. 
The $L_2$ barycenter is also a barycenter for the energy distance in 1D.


\section{CRPS and $L_2$-barycenter \label{app:crps_l2}}

Let $\mu_1,...,\mu_d$ be $d$ probability distributions, and $F_1,...,F_d$ their respective cumulative density functions. Their weighted-L2 barycenter is the probability distribution $\mu_{L_2}^{\mathbf{\lambda}}$ given by
\begin{align}
    \mu_{L_2}^{\mathbf{\lambda}} &= \sum_{i=1}^{n} \lambda_i \mu_i
\end{align}
where $\mathbf{\lambda}=(a_i,...\lambda_n)$ are the barycentric weights such that $\sum_{i=1}^{n}\lambda_i = 1$. 

Let $\nu_{obs}$ be the probability distribution of the truth or reference, and $F_{obs}$ its cumulative density function. In the case of a deterministic observation $y_r$, $\nu_{obs}=\delta_{y_{obs}}$ is a Dirac and $F_{obs}$ a step function.

The CRPS of two probability distributions is defined as the squared $L_2$-distance between their CDFs:
\begin{align}
    \text{CRPS}(\mu_i,\mu_j) &= \int_{-\infty}^{+\infty} \left[ F_i(u) - F_j(u) \right]^2 du
\end{align}

Thus, we have
\begin{align}
    \text{CRPS}(\mu_{L_2}^{\mathbf{\lambda}},\nu_{obs}) &= \int_{-\infty}^{+\infty} \left[ F_{L_2}^{\mathbf{\lambda}}(u) - F_{obs}(u) \right]^2 du \nonumber\\
        &= \int_{-\infty}^{+\infty} \left[  \sum_{i=1}^{n} \lambda_i F_i(u) - F_{obs}(u) \right]^2 du \nonumber\\
        &= \int_{-\infty}^{+\infty} \left[ \sum_{i=1}^{n} \lambda_i^2 F_i(u)^2 + F_{obs}(u)^2 +  \sum_{i=1}^{n} \sum_{j\neq i} \lambda_i \lambda_j F_i(u) F_j(u) - 2 \sum_{i=1}^{n} \lambda_i F_i(u) F_{obs}(u) \right] du \nonumber\\
        & \text{knowing that } \sum_{i=1}^{n}\lambda_i = 1 \Rightarrow \lambda_i = 1-\sum_{j\neq i}\lambda_j \Rightarrow \lambda_i^2 = \lambda_i (1 - \sum_{j\neq i}\lambda_j ) = \lambda_i - \lambda_i \sum_{j\neq i}\lambda_j \nonumber\\
        &=  \int_{-\infty}^{+\infty} \left[ \sum_{i=1}^{n} \lambda_i F_i(u)^2  + \sum_{i=1}^{n} \sum_{j\neq i} \lambda_i \lambda_j F_i(u)^2  + \sum_{i=1}^{n} \lambda_i F_{obs}(u)^2 \right. \nonumber\\
        & \quad\quad \left. + 2 \sum_{i=1}^{n} \sum_{j\geq i} \lambda_i \lambda_j F_i(u) F_j(u) - 2 \sum_{i=1}^{n} \lambda_i F_i(u) F_{obs}(u) \right] du \nonumber \\
        &= \int_{-\infty}^{+\infty} \left[ \sum_{i=1}^{n}\lambda_i[F_i(u)^2+F_{obs}(u)^2-2F_i(u)F_{obs}(u)] \right. \nonumber\\
        & \quad\quad \left. - \sum_{i=1}^{n} \sum_{j\geq i} \lambda_i \lambda_j [F_i(u)^2 + F_j(u)^2 - 2 F_i(u) F_j(u)] \right] du \nonumber \\
        &= \sum_{i=1}^{n}\lambda_i \left(\int_{-\infty}^{+\infty} \left[ F_i(u) - F_{obs}(u)\right]^2 du \right) - \sum_{i=1}^{n} \sum_{j\geq i} \lambda_i \lambda_j \left( \int_{-\infty}^{+\infty} \left[ F_i(u) - F_j(u)\right]^2 du \right) \nonumber \\
        &= \sum_{i=1}^{n}\lambda_i \text{CRPS}(\mu_i,\nu_{obs}) - \sum_{i=1}^{n} \sum_{j\geq i} \lambda_i \lambda_j \text{CRPS}(\mu_i,\mu_j) \nonumber \\
    \text{CRPS}(\mu_{L_2}^{\mathbf{\lambda}},\nu_{obs})    &= \sum_{i=1}^{n}\lambda_i \text{CRPS}(\mu_i,\nu_{obs}) - \sum_{i=1}^{n} \sum_{j\neq i} \frac{1}{2}\lambda_i \lambda_j \text{CRPS}(\mu_i,\mu_j) \label{eq:CRPS_l2}
\end{align}


\section{Spatial performance's significance \label{app:wilcoxon}}
\subsection{Barycenters with four models}

\begin{table}[ht]
\caption{Two-sided Wilcoxon test’s p-value for the 2m temperature (t2m) for weeks~3 and~4 combined. Bold indicates that the models are significantly different from each other at the $5\%$ significance level (i.e. p-values$<$ 0.05).}
\begin{tabular}{lllllll}
\tophline
\multicolumn{7}{c}{CRPSmean} \\ \middlehline
   &  ECMWF &  NCEP &  ECCC &  KMA &  $L_2$-bary &  $GaussW_2$-bary \\  \middlehline
 {$L_2$-bary} & { 3.5e-01 }& \textbf{ 3.5e-09 }& \textbf{ 3.3e-09 }& \textbf{ 5.8e-09 }& - & { 3.6e-01}  \\
  $GaussW_2$-bary & { 3.4e-01 }& \textbf{ 2.6e-09 }& \textbf{ 2.8e-09 }& \textbf{ 1.1e-08 }& { 3.6e-01 }& -  \\
\middlehline \middlehline
\multicolumn{7}{c}{CRPSp} \\ \middlehline
    &  ECMWF &  NCEP &  ECCC &  KMA &  $L_2$-bary &  $GaussW_2$-bary \\  \middlehline
  {$L_2$-bary} & { 9.0e-02 }& \textbf{ 4.4e-10 }& \textbf{ 2.0e-09 }& \textbf{ 1.7e-12 }& - & \textbf{ 6.5e-16}  \\
  $GaussW_2$-bary & \textbf{ 2.5e-03 }& \textbf{ 2.3e-12 }& \textbf{ 5.9e-12 }& \textbf{ 7.1e-15 }& \textbf{ 6.5e-16 }& -  \\
\middlehline \middlehline
\multicolumn{7}{c}{CRPSf} \\ \middlehline
    &  ECMWF &  NCEP &  ECCC &  KMA &  $L_2$-bary &  $GaussW_2$-bary \\  \middlehline
 {$L_2$-bary} & \textbf{ 1.1e-04 }& \textbf{ 6.9e-17 }& \textbf{ 1.8e-20 }& \textbf{ 2.0e-17 }& - & \textbf{ 3.9e-05}  \\
  $GaussW_2$-bary & \textbf{ 2.1e-05 }& \textbf{ 3.4e-18 }& \textbf{ 8.9e-21 }& \textbf{ 5.9e-18 }& \textbf{ 3.9e-05 }& -  \\
\bottomhline
\end{tabular}
\label{tab:sign_t2m}
\end{table}

\clearpage

\begin{table}[ht]
\caption{Two-sided Wilcoxon test’s p-value for the 10m wind speed (ws10m) for weeks~3 and~4 combined. Bold indicates that the models are significantly different from each other at the $5\%$ significance level (i.e. p-values$<$ 0.05).}
\begin{tabular}{lllllll}
\tophline
\multicolumn{7}{c}{CRPSmean} \\ \middlehline
    &  ECMWF &  NCEP &  ECCC &  KMA &  $L_2$-bary &  $GaussW_2$-bary \\ \middlehline
 {$L_2$-bary} & \textbf{ 2.5e-02 }& \textbf{ 1.2e-16 }& \textbf{ 6.8e-12 }& \textbf{ 6.7e-18 }& -  & { 7.2e-01 } \\
  $GaussW_2$-bary & \textbf{ 2.0e-02 }& \textbf{ 2.3e-16 }& \textbf{ 1.1e-11 }& \textbf{ 5.5e-18 }& { 7.2e-01 }& -  \\
\middlehline \middlehline
\multicolumn{7}{c}{CRPSp} \\ \middlehline
   &  ECMWF &  NCEP &  ECCC &  KMA &  $L_2$-bary &  $GaussW_2$-bary \\ \middlehline
 {$L_2$-bary} & \textbf{ 1.5e-02 }& \textbf{ 6.3e-12 }& \textbf{ 1.7e-11 }& \textbf{ 1.5e-14 }& - & \textbf{ 1.1e-22}  \\
  $GaussW_2$-bary & \textbf{ 1.3e-05 }& \textbf{ 9.7e-17 }& \textbf{ 3.6e-16 }& \textbf{ 1.2e-18 }& \textbf{ 1.1e-22 }& -  \\
\middlehline \middlehline
\multicolumn{7}{c}{CRPSf} \\ \middlehline
   &  ECMWF &  NCEP &  ECCC &  KMA &  $L_2$-bary &  $GaussW_2$-bary \\ \middlehline
 {$L_2$-bary} & \textbf{ 1.4e-19 }& \textbf{ 1.4e-26 }& \textbf{ 6.3e-31 }& \textbf{ 1.2e-26} & - & \textbf{ 2.1e-03}  \\
  $GaussW_2$-bary & \textbf{ 2.3e-20 }& \textbf{ 1.3e-26 }& \textbf{ 6.0e-31 }& \textbf{ 8.4e-27 }& \textbf{ 2.1e-03 }& -  \\
\bottomhline
\end{tabular}
\label{tab:sign_w10m}
\end{table}

\clearpage

\begin{table}[ht]
\caption{Two-sided Wilcoxon test’s p-value for the geopotential height (z500) for weeks~3 and~4 combined. Bold indicates that the models are significantly different  from each other at the $5\%$ significance level (i.e. p-values$<$ 0.05) with respect to the chosen score.}
\begin{tabular}{lllllll}
\tophline
\multicolumn{7}{c}{CRPSmean} \\  \middlehline
     &  ECMWF &  NCEP &  ECCC &  KMA &  $L_2$-bary &  $GaussW_2$-bary \\  \middlehline
 {$L_2$-bary} & \textbf{ 9.4e-03 }& \textbf{ 8.1e-08 }& \textbf{ 4.8e-04 }& \textbf{ 3.7e-06 }& - & { 3.0e-01}  \\
  $GaussW_2$-bary & \textbf{ 1.0e-02 }& \textbf{ 7.2e-08 }& \textbf{ 4.3e-04 }& \textbf{ 3.2e-06 }& { 3.0e-01 }& -  \\
\middlehline
\middlehline
 \multicolumn{7}{c}{CRPSp} \\  \middlehline 
   & ECMWF &  NCEP &  ECCC &  KMA &  $L_2$-bary &  $GaussW_2$-bary \\ \middlehline
{$L_2$-bary} & \textbf{ 1.3e-03 }& \textbf{ 3.6e-06 }& \textbf{ 8.4e-05 }& \textbf{ 1.1e-05} & - & \textbf{ 2.4e-12}  \\
  $GaussW_2$-bary & \textbf{ 3.0e-05 }& \textbf{ 9.4e-08 }& \textbf{ 1.6e-06 }& \textbf{ 1.2e-07 }& \textbf{ 2.4e-12 }& -  \\
\middlehline
\middlehline
\multicolumn{7}{c}{CRPSf} \\  \middlehline 
  &  ECMWF &  NCEP &  ECCC &  KMA &  $L_2$-bary &  $GaussW_2$-bary \\ \middlehline
 {$L_2$-bary} & \textbf{ 2.6e-09 }& \textbf{ 2.0e-13 }& \textbf{ 2.5e-17 }& \textbf{ 6.2e-08 }& - & { 9.8e-01 } \\
 $GaussW_2$-bary & \textbf{ 2.1e-09 }& \textbf{ 1.9e-13 }& \textbf{ 5.7e-17 }& \textbf{ 6.5e-08 }& { 9.8e-01 }& -  \\
\bottomhline
\end{tabular}
\label{tab:sign_z500}
\end{table}

\clearpage
\FloatBarrier

\subsection{Barycenters with three models \label{app:sign_3m}}

\begin{table}[ht!]
\small
\caption{Two-sided Wilcoxon test’s p-value for the 2m temperature (t2m) for weeks~3 and~4 combined. Bold indicates that the models are significantly different from each other at the $5\%$ significance level (i.e. p-values$<$ 0.05).}
\begin{tabular}{lllllllll}
\tophline
\multicolumn{9}{c}{CRPSmean} \\ \middlehline
    &   &  &   &  & \multicolumn{2}{c}{$L_2$-bary} &  \multicolumn{2}{c}{$GaussW_2$-bary} \\
      &  ECMWF &  NCEP &  ECCC &  KMA &  4 models &  3 models &   4 models &  3 models \\ \middlehline
 {$L_2$-bary \newline (3 models)} & { 6.4e-01 }& \textbf{ 1.4e-07 }& \textbf{ 4.4e-06 }& \textbf{ 3.3e-07 }& \textbf{ 4.1e-06 }& - & \textbf{ 5.6e-06 }& { 1.8e-01}  \\
  $GaussW_2$-bary  \newline (3 models) & { 5.8e-01 }& \textbf{ 1.5e-07 }& \textbf{ 5.7e-06 }& \textbf{ 4.5e-07 }& \textbf{ 6.8e-06 }& { 1.8e-01 }& \textbf{ 2.7e-06 }& -  \\
\middlehline \middlehline
\multicolumn{9}{c}{CRPSp} \\ \middlehline
    &   &  &   &  & \multicolumn{2}{c}{$L_2$-bary} &  \multicolumn{2}{c}{$GaussW_2$-bary} \\
     &  ECMWF &  NCEP &  ECCC &  KMA &  4 models &  3 models &   4 models &  3 models \\ 
 {$L_2$-bary \newline (3 models)} & { 8.9e-01 }& \textbf{ 3.3e-08 }& \textbf{ 1.4e-06 }& \textbf{ 2.0e-10 }& \textbf{ 3.6e-06 }& - & \textbf{ 5.4e-12 }& \textbf{ 3.4e-10}  \\
 $GaussW_2$-bary \newline (3 models) & { 2.7e-01 }& \textbf{ 9.3e-11 }& \textbf{ 1.5e-08 }& \textbf{ 1.8e-12 }& { 2.1e-01 }& \textbf{ 3.4e-10 }& \textbf{ 8.3e-07 }& -  \\
\middlehline \middlehline
\multicolumn{9}{c}{CRPSf} \\ \middlehline
    &   &  &   &  & \multicolumn{2}{c}{$L_2$-bary} &  \multicolumn{2}{c}{$GaussW_2$-bary} \\
      &  ECMWF &  NCEP &  ECCC &  KMA &  4 models &  3 models &   4 models &  3 models \\ 
{$L_2$-bary \newline (3 models)} & { 1.2e-01 }& \textbf{ 6.8e-13 }& \textbf{ 5.4e-17 }& \textbf{ 7.2e-16 }& \textbf{ 1.5e-07 }& - & \textbf{ 6.7e-10 }& \textbf{ 4.1e-02}  \\
  $GaussW_2$-bary  \newline(3 models) & { 7.6e-02 }& \textbf{ 6.4e-14 }& \textbf{ 5.3e-17 }& \textbf{ 2.7e-16 }& \textbf{ 9.7e-06 }& \textbf{ 4.1e-02 }& \textbf{ 2.5e-08 }& -  \\
\bottomhline
\end{tabular}
\label{tab:sign_t2m_3m}
\end{table}

\clearpage

\begin{table}[ht]
\caption{Two-sided Wilcoxon test’s p-value for the 10m wind speed (ws10m) for weeks~3 and~4 combined. Bold indicates that the models are significantly different from each other at the $5\%$ significance level (i.e. p-values$<$ 0.05).}
\begin{tabular}{lllllllll}
\tophline
\multicolumn{9}{c}{CRPSmean} \\ \middlehline
    &   &  &   &  & \multicolumn{2}{c}{$L_2$-bary} &  \multicolumn{2}{c}{$GaussW_2$-bary} \\
      &  ECMWF &  NCEP &  ECCC &  KMA &  4 models &  3 models &   4 models &  3 models \\ \middlehline
 {$L_2$-bary \newline (3 models)} & { 9.9e-01 }& \textbf{ 4.9e-13 }& \textbf{ 7.7e-09 }& \textbf{ 3.3e-15 }& \textbf{ 9.1e-10 }& - & \textbf{ 4.6e-10 }& { 4.8e-01 } \\
  $GaussW_2$-bary \newline (3 models) & { 8.2e-01 }& \textbf{ 5.9e-13 }& \textbf{ 9.2e-09 }& \textbf{ 2.0e-15 }& \textbf{ 2.1e-07 }& { 4.8e-01 }& \textbf{ 2.9e-09 }& -  \\
\middlehline \middlehline
\multicolumn{9}{c}{CRPSp} \\ \middlehline
    &   &  &   &  & \multicolumn{2}{c}{$L_2$-bary} &  \multicolumn{2}{c}{$GaussW_2$-bary} \\
      &  ECMWF &  NCEP &  ECCC &  KMA &  4 models &  3 models &   4 models &  3 models \\ 
 {$L_2$-bary \newline (3 models)} & { 6.1e-01 }& \textbf{ 3.1e-09 }& \textbf{ 5.8e-08 }& \textbf{ 1.3e-12 }& \textbf{ 3.1e-06 }& - & \textbf{ 9.7e-17 }& \textbf{ 4.5e-20}  \\
  $GaussW_2$-bary \newline (3 models) & { 5.1e-02 }& \textbf{ 6.4e-14 }& \textbf{ 1.6e-11 }& \textbf{ 2.6e-17 }& { 8.9e-01 }& \textbf{ 4.5e-20 }& \textbf{ 1.1e-06 }& -  \\
\middlehline \middlehline
\multicolumn{9}{c}{CRPSf} \\ \middlehline
    &   &  &   &  & \multicolumn{2}{c}{$L_2$-bary} &  \multicolumn{2}{c}{$GaussW_2$-bary} \\
      &  ECMWF &  NCEP &  ECCC &  KMA &  4 models &  3 models &   4 models &  3 models \\ 
 {$L_2$-bary \newline (3 models)} & \textbf{ 3.3e-09 }& \textbf{ 6.9e-22 }& \textbf{ 8.8e-30 }& \textbf{ 1.4e-25 }& \textbf{ 2.3e-11 }& - & \textbf{ 4.2e-13 }& \textbf{ 3.8e-02}  \\
  $GaussW_2$-bary \newline (3 models)& \textbf{ 4.1e-10 }& \textbf{ 4.3e-22 }& \textbf{ 6.2e-30 }& \textbf{ 7.7e-26 }& \textbf{ 1.7e-08 }& \textbf{ 3.8e-02 }& \textbf{ 2.1e-11 }& -  \\
\bottomhline
\end{tabular}
\label{tab:sign_w10m_3m}
\end{table}

\clearpage

\begin{table}[ht]
\caption{Two-sided Wilcoxon test’s p-value for the geopotential height (z500) for weeks~3 and~4 combined. Bold indicates that the models are significantly different from each other at the $5\%$ significance level (i.e. p-values$<$ 0.05).}
\begin{tabular}{lllllllll}
\tophline
\multicolumn{9}{c}{CRPSmean} \\ \middlehline
    &   &  &   &  & \multicolumn{2}{c}{$L_2$-bary} &  \multicolumn{2}{c}{$GaussW_2$-bary} \\
      &  ECMWF &  NCEP &  ECCC &  KMA &  4 models &  3 models &   4 models &  3 models \\ 
 \middlehline
 {$L_2$-bary \newline (3 models)} & { 2.0e-01 }& \textbf{ 7.9e-07 }& \textbf{ 3.3e-03 }& \textbf{ 2.1e-05 }& \textbf{ 1.3e-02 }& - & \textbf{ 1.0e-02 }& { 2.4e-01}  \\
  $GaussW_2$-bary \newline (3 models)& { 1.9e-01 }& \textbf{ 8.9e-07 }& \textbf{ 2.8e-03 }& \textbf{ 2.4e-05 }& \textbf{ 2.0e-02 }& { 2.4e-01 }& \textbf{ 9.2e-03 }& -  \\
\middlehline \middlehline
\multicolumn{9}{c}{CRPSp} \\ \middlehline
   &   &  &   &  & \multicolumn{2}{c}{$L_2$-bary} &  \multicolumn{2}{c}{$GaussW_2$-bary} \\
     &  ECMWF &  NCEP &  ECCC &  KMA &  4 models &  3 models &   4 models &  3 models \\
 {$L_2$-bary \newline (3 models)} & { 7.2e-02 }& \textbf{ 4.9e-06 }& \textbf{ 6.5e-04 }& \textbf{ 8.8e-05 }& { 7.9e-02 }& - & \textbf{ 5.9e-05 }& \textbf{ 2.8e-07}  \\
  $GaussW_2$-bary \newline (3 models) & \textbf{ 1.8e-02 }& \textbf{ 3.7e-07 }& \textbf{ 4.3e-05 }& \textbf{ 2.9e-06 }& { 5.7e-01 }& \textbf{ 2.8e-07 }& \textbf{ 3.1e-02 }& -  \\
\middlehline \middlehline
\multicolumn{9}{c}{CRPSf} \\ \middlehline
   &   &  &   &  & \multicolumn{2}{c}{$L_2$-bary} &  \multicolumn{2}{c}{$GaussW_2$-bary} \\
      &  ECMWF &  NCEP &  ECCC &  KMA &  4 models &  3 models &   4 models &  3 models \\
 {$L_2$-bary \newline (3 models)} & \textbf{ 4.3e-05 }& \textbf{ 9.2e-13 }& \textbf{ 1.0e-13 }& \textbf{ 8.6e-06 }& \textbf{ 5.4e-03 }& - & \textbf{ 2.3e-03 }& { 8.3e-01 } \\
  $GaussW_2$-bary  \newline (3 models) & \textbf{ 2.5e-05 }& \textbf{ 5.9e-13 }& \textbf{ 5.5e-14 }& \textbf{ 9.2e-06 }& \textbf{ 1.3e-02 }& { 8.3e-01 }& \textbf{ 3.2e-03 }& -  \\
\bottomhline
\end{tabular}
\label{tab:sign_z500_3m}
\end{table}

\FloatBarrier


\end{document}